\documentclass[aps,twocolumn,superscriptaddress,prd]{revtex4-1}

\usepackage[T1]{fontenc}
\usepackage{times}
\usepackage{euscript}
\usepackage{amsthm}
\usepackage{graphicx}
\usepackage{amsmath}
\usepackage{amssymb}
\usepackage{amsfonts}
\usepackage{epstopdf}
\usepackage{multirow,makecell,array}
\usepackage{mathtools}
\usepackage{epstopdf}
\usepackage{color}
\usepackage{bbm}
\usepackage{bbold}

\usepackage{comment}

\allowdisplaybreaks

\newcommand*{\I}{ {\rm i} }
\newcommand*{\ee}{ {\rm e} }

\definecolor{orange}{RGB}{255,102,0}

\begin{document}
\thispagestyle{empty}

\title{The Heisenberg-Wigner formalism for transverse fields}
\author{Christian Kohlf\"urst}\email{c.kohlfuerst@hzdr.de}
\affiliation{Helmholtz-Zentrum Dresden Rossendorf e.V., Bautzner Landstraße 400, D-01328 Dresden, Germany}

\begin{abstract}
  We discuss the Heisenberg-Wigner phase-space formalism in quantum electrodynamics as well as scalar quantum electrodynamics with respect to transverse fields. In regard to the special characteristics of such field types we derive modified transport equations such that particle momenta perpendicular to the propagation direction of the waves show up as external parameters only.
  In case of spatially oscillating fields we further demonstrate how to transform momentum derivative operators of infinite order into simple coupling terms.
\end{abstract}

\maketitle

\section{Introduction}
Strong-field quantum electrodynamics (SFQED) unites special relativity, electrodynamics and quantum physics within one formalism \cite{euler_heisenberg}. As such it is capable of describing fundamental processes in physics to an astonishing level of precision, see e.g. lamb shift \cite{Lamb} or the magnetic moment of the electron \cite{SchwingerMu}.
Additionally, the theory stimulated predictions on the fabric of time and space itself including the possibility of a decaying vacuum \cite{schwinger_1951}.

One of the most striking examples of such strong-field effects is the conversion of energy into massive particles either through the interaction of multiple high-energy photons \cite{BreitWheeler, Narozhnyi, Brezin:1970xf, Reiss}, see also Refs.~\cite{PhysRevD.94.013010,PhysRevD.105.L071902,Mercuri-Baron:2021waq,Blackburn:2021cuq, Salgado:2021uua} on recent developments, or by having particles come into existence through stimulation by a high-intensity field \cite{sauter_1931}. Such an effect would be impossible in the context of Maxwell's equation where light does not interact with light. Quantum electrodynamics changes this picture completely, as at high intensities light-by-light scattering enabled through virtual particles acting as mediators becomes possible \cite{euler_heisenberg, Weisskopf}. 

To probe the scattering of light with light in detail in an earth-based laboratory effectively putting these non-linear interaction terms at display poises strong-field experiments to become one of the most anticipated experiments at modern laser facilities \cite{Marklund:2008gj, Dunne:2008kc, Heinzl:2008an, dipiazza_rmp_2012, Gies:2008wv, PhysRevLett.129.061802}. Prominent examples in the context of pair production are the SLAC e-340 at FACET-II at the linear collider in Stanford (LCLS) \cite{FACET-II} or the LUXE experiment at DESY in Hamburg \cite{LUXE}. Both of which will probe strong-field physics with unprecedented precision. For reference, a proof of principle experiment within SLAC e-144 \cite{Burke:1997ewA, Burke:1997ewB}.

In this regard, experiments in the regime of quantum plasmas or even beyond require a solid theoretical description in order to make a comparison between theoretical predictions and experimental data viable. As such, a variety of approaches have been introduced in the past; worldline instantons \cite{ SemiClassA, PhysRevD.73.065028, Affleck, PhysRevD.84.125023}, Monte-Carlo worldline methods \cite{Gies:2005bz}, directly solving the Dirac equation \cite{Aleksandrov:2016lxd, Ruf, PhysRevA.97.022515, PhysRevA.81.022122}, quantum kinetic theories \cite{Smolyansky:1997fcC,Vasak:1987umA,Kluger}, WKB approaches \cite{linder_prd_2015,Kim:2007pm,Oertel,Taya}, real-time lattice techniques \cite{Hebenstreit:2013baa, Hebenstreit:2013qxa}, techniques that embrace analogies \cite{Gies:2015hia} and even tools that benefit from a hybrid approach mixing semi-classical and computational methods \cite{li_prd_2017, Blinne:2015zpa,NewPaper}. 
While undoubtedly important in gaining new knowledge, most of these methods are only applicable in specific regions of the full parameter space. For example, direct $n$-photon scatterings (at comparably low intensities) can be described perturbatively while for strong, slowly varying fields an analysis based on the Euler-Heisenberg Lagrangian yields good results \cite{euler_heisenberg, Kohlfurst:2017git}. The crux is the intermediate region where no obvious expansion parameter exists and different effects contribute equally, see Refs. \cite{Dittrich:2000zu, Dunne:2012vv, Gelis:2015kya, Fedotov} for in-depth reviews regarding our theoretical understanding of strong-field QED in general.
  
One approach that can probe different regimes of pair production is the Heisenberg-Wigner formalism \cite{Vasak:1987umA, NewPaper}, which takes into account the possibility of creating particle pairs at intermediate times \cite{PhysRevD.83.025011, PhysRevD.94.065005, PhysRevD.105.016021, Diez}. Additionally, the general particle dynamics is incorporated to all orders of $\hbar$ thus including not only the Lorentz force or the Stern-Gerlach force but also spin dynamics even beyond the BMT equations \cite{BB}. 

Over the last years there was steady progress not only in understanding the intricacies of the derived transport equations, see the theses, Refs. \cite{Hebenstreit:2011pm, KohlfurstDiss, KohlfurstMag, DiezMag}, but also in finding and applying computational methods that render solving these equations feasible \cite{Hebenstreit, KohlfurstTech}. Within this manuscript we take another step in the direction of creating a system of equations that are easy-to-solve while maintaining all aspects of pair production in strong background fields.

As we want to deliver a complete picture of pair production for the specific class of transverse fields, we will derive the quantum kinetic transport equations for QED as well as scalar QED, see. Secs. \ref{sec:dhw} and \ref{sec:fvhw}. In Sec. \ref{sec:transverse} we aim at providing general information on the structure of the transport equations in transverse fields. Most importantly, we derive the pseudo-differential operators incorporating the various features that make transverse fields so uniquely well suited for application within the Heisenberg-Wigner formalism. In the third section, Sec. \ref{sec:Appl}, we discuss an exemplary field configuration on the basis of the Heisenberg-Wigner formalism. Specifically we display the transport equations for bi-frequent field configurations in the context of Dirac particles in Sec. \ref{sec:Appl_DHW} and for scalar particles in Sec. \ref{sec:Appl_FVHW}. 
At the end we provide a brief conclusion regarding the main points of the manuscript, Sec. \ref{sec:Conclusion}, and state our opinion on possible further applications and future pathways of the Heisenberg-Wigner formalism, Sec. \ref{sec:Outlook}.

Throughout the manuscript we use natural units $c=\hbar = 1$.
 
\section{Dirac-Heisenberg-Wigner formalism}\label{sec:dhw}%
 
The Heisenberg-Wigner formalism is an approach to describe quantum physics in phase-space similar to classical kinetic theories. The main conceptual difference between classical and quantum systems is the uncertainty principle which prevents an interpretation in terms of particles or particle distributions in phase-space. Instead, the phase-space of quantum systems deals in terms of quasi-probabilities which can easily become negative. Only when these issues with a negative particle number are addressed, for example, by integrating out either momentum or spatial coordinates or by performing a convolution with a smearing function, an interpretation in terms of distribution functions is valid \cite{ShinRafelski}. 

Despite these shortcomings, quantum phase-space formalisms provide an incredibly detailed insight into non-equilibrium statistical systems that are governed by principles of the quantum world. Incorporating Dirac (anti)-spinors and electromagnetic fields on the basis of the QED Lagrangian into the Heisenberg-Wigner formalism leads, for example, to a powerful relativistic, quantum kinetic theory capable of describing the interaction of photons with spin-$1/2$ fermions \cite{Vasak:1987umA, BB, PhysRevLett.98.025001, PhysRevE.104.015207, PhysRevE.102.043203, PhysRevE.100.023201}. 

To expand on this point in more detail, the Dirac-Heisenberg-Wigner (DHW) formalism gives rise to transport equations that determine the time evolution of the underlying distribution functions, e.g., the particle number distribution. In this way, there is no need to keep track of all individual particles as instead of the dynamics of a wave packet representing one single particle the time-evolution of a macroscopic average over an ensemble of particles is considered. 

An additional advantage of the Dirac-Heisenberg-Wigner formalism is that it not only takes into account the particle dynamics in an external field, it also grants access to the particles' charge distribution or the spin density distribution. Moreover, the conversion from photons to pairs of electrons and positrons is deeply integrated into the formalism thus no assumptions on particle creation rates have to be artificially implemented. In this way, it is possible to employ phase-space methods to actively search for unknown effects or even emergent phenomena in non-equilibrium quantum plasmas.

The downside of such an approach is its huge computational cost. While the formalism grants access to a variety of important quantities, meaningful results can only be achieved if the corresponding phase-space shows a resolution small enough to capture important quantum statistical details. Furthermore, the time-integration of the quantum system has to be evaluated accurately. Considering an, in general, $2n$-dimensional phase-space, with $n$ being the dimension of the system, solving the full transport equations remains a massive undertaking even for modern computers.

Nevertheless, in order to have a self-consistent manuscript we state the key points in the original derivation of the quantum transport equations in the following. We will, however, not go too much into detail about the significance of each element in the derivation. We recommend Ref. \cite{Vasak:1987umA} for an in-depth analysis of the transport equations. In Refs. \cite{BB,Hebenstreit,KohlfurstDiss,Vasak:1987umB} the further development of the formalism is laid out. Additional information regarding the Wigner function formalism can be found in Refs. \cite{Weinbub, Ochs:1998qj, Hidaka:2022dmn}. 

The basis of any Heisenberg-Wigner formalism is given by the density operator
\begin{equation}
 \hat {\mathcal C}_{\alpha \beta} \left( r , s \right) = \mathcal U \left(A,r,s \right) \left[ \bar {\Psi}_\beta \left( r - s/2 \right), {\Psi}_\alpha \left( r + 
s/2 \right) \right], \label{equ:C}
\end{equation}
with the center-of-mass coordinate $r$ and the relative coordinate $s$. As we consider the electromagnetic field strength tensor $F_{\mu\nu}$, and thus the vector potential $A_{\mu}$, as a c-number instead of as an operator no path ordering is required \cite{Vasak:1987umA}. Instead, we ensure gauge-invariance through implementation of a Wilson line factor
\begin{equation}
 \mathcal U \left(A,r,s \right) = \exp \left( \mathrm{ie} \int {\rm d} 
\xi \ A \left(r+ \xi s \right) \, s \right). \label{equ:U}
\end{equation}
Here, the spinors $\Psi$ are determined by the Lagrangian for quantum electrodynamics describing charged spin-$1/2$ particles   
\begin{multline}
{\mathcal L} \left( \Psi, \bar{\Psi}, A \right) = \\
\frac{1}{2} \left( {\rm i} \bar{\Psi} \gamma^{\mu} \mathcal{D}_{\mu} \Psi - {\rm i} \bar{\Psi} \mathcal{D}_{\mu}^{\dag} \gamma^{\mu} \Psi \right) 
 -m \bar{\Psi} \Psi - \frac{1}{4} F_{\mu \nu} F^{\mu \nu}. \label{equ:Lag}
\end{multline}
In Eq. \eqref{equ:Lag}, the quantity $m$ determines the masses of particle and anti-particle in the formalism (here, electrons and positrons), $\mathcal{D}_{\mu} =  \partial_{\mu} +{\rm i} e A_{\mu} $ and $\mathcal{D}_{\mu}^{\dag} = \overset{\leftharpoonup} {\partial_{\mu}} -{\rm i} e A_{\mu} $ describe covariant derivatives and $\gamma^\mu$ are the Dirac matrices.

In order to obtain a kinetic formalism in familiar position- and momentum coordinates, a Fourier transform in $s$ is to be performed. As a result we obtain the covariant Wigner operator
\begin{align}
 \hat{\mathcal W}_{\alpha \beta} \left( r , p \right) = \frac{1}{2} \int {\rm d}^4 s \ \mathrm{e}^{\mathrm{i} ps} \  \hat{\mathcal C}_{\alpha \beta} \left( r , s \right). \label{equ:W}
\end{align}
On the basis of the Lagrangian \eqref{equ:Lag} we obtain the (adjoint) Dirac equations determining the equations of motion for the spinors $\Psi,~ {\bar \Psi}$. The Wigner operator \eqref{equ:W} follows the same principles as the spinors it is based on, thus we can rewrite the (adjoint) Dirac equation leading to  
\begin{alignat}{3}
 & \left( \frac{1}{2} \hat D_{\mu}  - {\rm i} \hat P_{\mu}  \right) \gamma^{\mu} \hat{\mathcal W} \left( r , p \right) && = - &&{\rm i} m \hat{\mathcal W} \left( r , p \right), \label{equ:W1} \\
 & \left( \frac{1}{2} \hat D_{\mu}  + {\rm i} \hat P_{\mu}  \right) \hat{\mathcal W} \left( r , p \right) \gamma^{\mu}  && = &&{\rm i} m \hat{\mathcal W} \left( r , p \right). \label{equ:W2} 
\end{alignat}
In this regard, the covariant derivatives $\mathcal{D}_{\mu}$ and $\mathcal{D}_{\mu}^{\dag}$ are replaced by nonlocal, pseudo-differential operators 
\begin{alignat}{4}
 & \hat D_{\mu}  && = \partial_{\mu}^r - e &&\int_{-1/2}^{1/2} {\rm d} \xi \ && F_{\mu \nu} \left( r - {\rm i} \xi \partial^p \right) \partial_p^{\nu}, \label{equ:D} \\
 & \hat P_{\mu}  && = p_{\mu} - {\rm i} e && \int_{-1/2}^{1/2} {\rm d} \xi \ \xi \ && F_{\mu \nu} \left( r - {\rm i} \xi \partial^p \right) \partial_p^{\nu}, \label{equ:P}
\end{alignat}
having the electromagnetic field strength tensor $F_{\mu \nu}$ replace the vector potential $A_{\mu}$ and formally featuring derivative operators $\partial^p$ as arguments. \cite{footnote1}. The integration path is chosen such that $p$ can be identified as the kinetic momentum.

An equation of motion for the matrix-valued Wigner function $\mathcal W$, the vacuum expectation value of the Wigner operator
\begin{equation}
 \mathcal W \left( r , p \right) = \langle \Phi | \hat{\mathcal W} \left( r , p \right) | \Phi \rangle,
\end{equation}
is obtained by taking the vacuum expectation values of Eqs. \eqref{equ:W1} and \eqref{equ:W2}. This step is crucial in order to have access to a system of transport equations in terms of distribution functions. Note, that at this point 
the consequences of treating the background fields as classical c-number fields instead of operators becomes obvious. While terms of the form $\langle \Omega | \hat{\mathcal W} \left( r , p \right) \ \hat F^{\mu \nu} (r) | \Omega \rangle$ would create an infinite chain of coupled equations (BBGKY hierarchy, see, e.g., Ref. \cite{Fauth}), a Hartree-type approximation as is used in this manuscript yields a truncation at first order \cite{Ochs:1998qj}.

While such measures exempts us from describing, e.g., radiative emission, it marks an important step towards constraining the total number of interactions and hence differential equations to a computationally manageable level. Besides, our primary interest in this manuscript are fundamental particle production rates in subcritical fields, which can still be obtained despite the decision to use c-number fields. 

To keep the notation clear and simple, we expand the matrix-valued function $\mathcal W$ into Dirac bilinears 
\begin{multline}
\mathcal W  \left( r , p \right) = \\ 
\frac{1}{4} \left( \mathbbm{1} \mathbbm{S} + {\rm i} \gamma_5 \mathbbm{P} + \gamma^{\mu} \mathbbm{V}_{\mu} + \gamma^{\mu} \gamma_5 \mathbbm{A}_{\mu} + \sigma^{\mu \nu} \mathbbm{T}_{\mu \nu} \right). \label{equ:wigner}
\end{multline}
This step allows us to determine the time evolution of each individual component of the matrix creating an easier-to-interpret system of equations. In this regard, we also have to abandon covariance as it demands a description at all times. As such a formulation is too restrictive for our needs, we project on equal times, $\int {\rm d}p_0 / (2 \pi)$, effectively transforming the equations of motion into an initial-value problem governed by transport equations in the equal-time Wigner components
${\mathbbm w} \left( t, \boldsymbol{x} , \boldsymbol{p} \right) = \int {\rm d}p_0 / (2 \pi) \ \mathbbm{W} \left( r , p \right)$. Ultimately, the equations of motion for the individual Wigner components read
 \begin{alignat}{4}
    & D_t \mathbbm{s}     && && -2 \boldsymbol{\Pi} \cdot \mathbbm{t_1} &&= 0, \label{eq_DHW1} \\
    & D_t \mathbbm{p} && && +2 \boldsymbol{\Pi} \cdot \mathbbm{t_2} &&= -2m\mathbbm{a}_\mathbb{0},  \\
    & D_t \mathbbm{v}_\mathbb{0} &&+ \boldsymbol{D} \cdot \mathbbm{v} && &&= 0, \\
    & D_t \mathbbm{a}_\mathbb{0} &&+ \boldsymbol{D} \cdot \mathbbm{a} && &&=+ 2m\mathbbm{p}, \\    
    & D_t \mathbbm{v} &&+ \boldsymbol{D} \ \mathbbm{v}_\mathbb{0} && +2 \boldsymbol{\Pi} \times \mathbbm{a} &&= -2m\mathbbm{t_1}, \\    
    & D_t \mathbbm{a} &&+ \boldsymbol{D} \ \mathbbm{a}_\mathbb{0} && +2 \boldsymbol{\Pi} \times \mathbbm{v} &&= 0, \\
    & D_t \mathbbm{t_1} &&+ \boldsymbol{D} \times \mathbbm{t_2} && +2 \boldsymbol{\Pi} \ \mathbbm{s} &&= +2m\mathbbm{v}, \\    
    & D_t \mathbbm{t_2} &&- \boldsymbol{D} \times \mathbbm{t_1} && -2 \boldsymbol{\Pi} \ \mathbbm{p} &&= 0.  \label{eq_DHW2} 
  \end{alignat} 
with $\mathbbm{t_1} = 2 \mathbbm{t}^{i0} \boldsymbol{e}_i$ and $\mathbbm{t_2} = \epsilon_{ijk} \mathbbm{t}^{jk} \boldsymbol{e}_i$. The corresponding pseudo-differential operators $D_t ,~ \boldsymbol{D}$ and $\boldsymbol{\Pi}$ are given by
  \begin{alignat}{6}
     & D_t && = \partial_t &&+ e &&\int {\rm d} \xi &&\boldsymbol{E} \left( \boldsymbol{x}+{\rm i} \xi \boldsymbol{\nabla}_p,t \right) && ~ \cdot \boldsymbol{\nabla}_p, \label{eqn2_1}  \\[-2mm]
     & \boldsymbol{D} && = \boldsymbol{\nabla}_x &&+ e &&\int {\rm d} \xi &&\boldsymbol{B} \left( \boldsymbol{x}+{\rm i} \xi \boldsymbol{\nabla}_p,t \right) &&\times \boldsymbol{\nabla}_p,  \label{eqn2_2} \\[-2mm]
     & \boldsymbol{\Pi} && = \boldsymbol{p} &&- {\rm i} e &&\int {\rm d} \xi \xi &&\boldsymbol{B} \left( \boldsymbol{x}+{\rm i} \xi \boldsymbol{\nabla}_p,t \right) &&\times \boldsymbol{\nabla}_p.  \label{eqn2_3}
  \end{alignat}     
Vacuum initial conditions are given by  
\begin{alignat}{3}
\mathbbm{s}_{\rm vac} \left(\boldsymbol{p} \right) = -\frac{2m}{\sqrt{m^2 +
\boldsymbol{p}^2}}, \, \, 
  \boldsymbol{\mathbbm{v}}_{\rm vac} \left(\boldsymbol{p} \right) = -\frac{2
\boldsymbol{p}}{\sqrt{m^2 + \boldsymbol{p}^2}}, \label{equ:vac}
\end{alignat}
with all other Wigner components vanishing initially \cite{BB, Hebenstreit:2011pm}.

Although the individual Wigner components are not directly observable an interpretation in terms of familiar quantities is possible. As such, we have the mass density $ \mathbbm{s}$, the charge density $ \mathbbm{v}_\mathbb{0}$, the current density $ \boldsymbol{\mathbbm {v}}$, the spin density $ \boldsymbol{\mathbbm {a}}$ and the magnetic moment density $ \boldsymbol{\mathbbm {t_2}}$ \cite{BB}. 

Furthermore, Noether's theorem can be applied yielding a prescription on how to obtain observables at asymptotic times $t \to \infty$, e.g., the particle distribution function \cite{Hebenstreit:2011pm}
\begin{equation}
n \left( \boldsymbol{x}, \boldsymbol{p} \right) = \frac{m \left( \mathbbm{s}-\mathbbm{s}_{\rm vac} \right) + {\boldsymbol p} \cdot \left(
\boldsymbol{\mathbbm{v}}-\boldsymbol{\mathbbm{v}}_{\rm vac} \right)}{2\sqrt{m^2+\boldsymbol{p}^2}}
 \label{equ:n}
\end{equation}
or the particle momentum spectrum
\begin{equation}
 n \left( \boldsymbol{p} \right) = \int \frac{{\rm d}^3 {\boldsymbol x}}{\left(2 \pi \right)^3} ~ n \left( \boldsymbol{x}, \boldsymbol{p} \right).
 \label{equ:nn}
\end{equation}

\section{Feshbach-Villars-Heisenberg-Wigner formalism}
\label{sec:fvhw}

To put the versatility of the Heisenberg-Wigner formalism on display we will also consider scalar quantum electrodynamics (sQED) which deals with charged particles of spin zero. In this regard, sQED is a simpler theory as there is no need to track the spin degrees of freedom or consider light-matter interactions that are facilitated by the particle spin.  


Additionally, a formalism based on bosons provides us with another opportunity to dissect the intricacies of pair production. 
For example, it has been shown that in order to obtain the angular momentum distribution, which is the gateway to fully understanding the momentum spectrum, a two-step calculation process might be the favorable strategy \cite{KohlfurstSpin}. As within the DHW formalism one can hardly distinguish between different electron-positron spin states a dual calculation taking into account spin-zero pairs gives the opportunity to distinguish ortho- and para-states. This is an important distinction as depending on the coupling of the particle spins with the background field a particle pair is created with a particular probability and a particular orbital angular momentum thus decisively changing the particle spectrum \cite{SeiptKing, Seipt:2020uxv}. 

Similar to the overview we have given regarding the Dirac-Heisenberg-Wigner formalism, Sec. \ref{sec:dhw}, we also state the key elements in the Feshbach-Villars-Heisenberg-Wigner formalism culminating in the transport equations for particles of spin zero, see Refs. \cite{ZhuangHeinz, FVHW, BestGreiner} for detailed discussions on the subject. 

We state the Lagrangian 
\begin{equation}
\mathcal{L}_{\rm sQED} \left( \phi, F \right) = \frac{1}{2} (D_\mu \phi) D^\mu \phi - \frac{m^2}{2} \phi^2 -\frac{1}{4} F_{\mu\nu}F^{\mu\nu}\ , \label{equ:Lag_scal}
\end{equation}
with the scalar field $\phi$ as well as the covariant derivative $\mathcal{D}_{\mu} = \partial_{\mu} +\I e A_{\mu}$ and the vector potential $A_\mu$  connected to the field strength tensor through $F_{\mu\nu} = \partial_\mu A_\nu - \partial_\nu A_\mu$. On the basis of Eq. \eqref{equ:Lag_scal} we obtain the Klein-Gordon equation
\begin{eqnarray}
\left( D_\mu D^\mu +m^2 \right) \phi = 0, \label{equ:KG}
\end{eqnarray}
defining the equation of motion for a scalar field. While the Heisenberg-Wigner formalism can also be applied on $\phi$ with respect to Eq. \eqref{equ:KG} the presence of second order derivatives in time is expected to create a challenging problem \cite{ZhuangHeinz}. Hence, at this point we switch to a two-component Feshbach-Villars field description $\Phi = \begin{pmatrix} \xi \\ \chi \end{pmatrix}$,
\begin{align}
 \xi &= \frac{1}{2} \left( \phi + \frac{\I}{m} \partial_t \phi - \frac{e A_0}{m} \phi \right), \\
 \chi &= \frac{1}{2} \left( \phi - \frac{\I}{m} \partial_t \phi + \frac{e A_0}{m} \phi \right).
\end{align}
In this way, the equation of motion is given  in terms of a first-order differential equation in time
\begin{equation}
 \I \partial_t \Phi = \left( \frac{1}{2m} \left(-\I \boldsymbol{\nabla} -e \boldsymbol{A} \right)^2 \left(\sigma_3 + \I \sigma_2 \right) + m \sigma_3 + e A_0 \right) \Phi, \label{equ:FVKG}
\end{equation}
where $\sigma_1$, $\sigma_2$ and $\sigma_3$ are the Pauli matrices.
A further convenience is that the field $\Phi$ plays the same role as the spinor $\Psi$ did for the Dirac-Heisenberg-Wigner formalism making the derivation of the transport equations strikingly similar.
\cite{footnote2}

The basic density operator is thus given here in terms of Feshbach-Villars fields
\begin{multline}
\hat {\mathcal C}^{\rm FV}_{\alpha \beta} \left( t, \boldsymbol{x} , \boldsymbol{s} \right) = \mathcal U^{\rm FV} \left(\boldsymbol{A},t, \boldsymbol{x}, \boldsymbol{s} \right) \\
\times \left[ {\Phi}^\dag_\beta \left( \boldsymbol{x} - \boldsymbol{s}/2,t \right), {\Phi}_\alpha \left( \boldsymbol{x} + 
\boldsymbol{s}/2,t \right) \right], 
\end{multline}
where we have the Wilson line factor
\begin{multline}
 \mathcal U^{\rm FV} \left(\boldsymbol{A},t,\boldsymbol{x},\boldsymbol{s} \right) = \\
 \exp \left(- \mathrm{i}e \int_{-1/2}^{1/2} {\rm d} 
\xi \ \boldsymbol{A} \left(\boldsymbol{x}+ \xi \boldsymbol{s},t \right) \cdot \boldsymbol{s} \right), \label{eq:FV_W}
\end{multline}
with vector potential $\boldsymbol{A}$ for center-of-mass $r=\left(t, \boldsymbol{x} \right)$ and relative coordinates $\boldsymbol{s}$. Again, the background fields are considered to be c-numbers thus no path-ordering is needed. The Wilson line \eqref{eq:FV_W} has the same function as for the DHW formalism simply ensuring gauge-invariance. 

Performing the Fourier transform of the density operator $\hat {\mathcal C}^{\rm FV} \left( t, \boldsymbol{x}, \boldsymbol{s} \right)$ we directly obtain the equal-time Wigner operator
\begin{equation}
 \hat {\mathcal W}^{\rm FV}_{\alpha \beta} \left( t, \boldsymbol{x}, \boldsymbol{p} \right) = \frac{1}{2} \int {\rm d}^3 s \ \mathrm{e}^{-\mathrm{i} \boldsymbol{p} \cdot  \boldsymbol{s}} \  \hat{\mathcal C}^{\rm FV}_{\alpha \beta} \left( t, \boldsymbol{x}, \boldsymbol{s} \right), \label{equ:Wf}
\end{equation}
with momentum $\boldsymbol{p}$. The equations of motion for the operator $\hat{\mathcal W}^{\rm FV}$ are then directly determined by the field equation \eqref{equ:FVKG} and its adjoint. Furthermore, we take the vacuum expectation value of the equation of motion, thus determing the time evolution of the equal-time Wigner function
\begin{equation}
 {\mathcal W}^{\rm FV}_{\alpha \beta} \left( t, \boldsymbol{x}, \boldsymbol{p} \right) = \langle \Omega | \hat {\mathcal W}^{\rm FV}_{\alpha \beta} \left( t, \boldsymbol{x}, \boldsymbol{p} \right) | \Omega \rangle. \label{equ:WFunc}
\end{equation}
The corresponding equation reads
\begin{multline}
2m D_t {\mathcal W}^{\rm FV} \\
+\I \left( \frac{1}{4} \boldsymbol{D}^2 - \boldsymbol{\Pi}^2 \right) \left( {\mathcal W}^{\rm FV} \left( \sigma_3 - \I \sigma_2 \right) - \left( \sigma_3 + \I \sigma_2 \right) {\mathcal W}^{\rm FV} \right) \\
+ \boldsymbol{\Pi} \cdot \boldsymbol{D} \left( {\mathcal W}^{\rm FV} \left( \sigma_3 - \I \sigma_2 \right) - \left( \sigma_3 + \I \sigma_2 \right) {\mathcal W}^{\rm FV} \right) \\
-2 \I m^2 \left( 
{\mathcal W}^{\rm FV} \sigma_3 - \sigma_3 {\mathcal W}^{\rm FV} \right) = 0, \label{equ:WF} 
\end{multline}
where we find the familiar, nonlocal, pseudo-differential operators \eqref{eqn2_1}-\eqref{eqn2_3} 
\begin{alignat}{6}
  & D_t && = \partial_t &&+ e &&\int {\rm d} \xi &&\boldsymbol{E} \left( \boldsymbol{x}+\I \xi \boldsymbol{\nabla}_p,t \right) && ~ \cdot \boldsymbol{\nabla}_p, \label{eq_FVHW_Der1} \\
  & \boldsymbol{D} && = \boldsymbol{\nabla}_x &&+ e &&\int {\rm d} \xi &&\boldsymbol{B} \left( \boldsymbol{x}+\I \xi \boldsymbol{\nabla}_p,t \right) &&\times \boldsymbol{\nabla}_p, \\
  & \boldsymbol{\Pi} && = \boldsymbol{p} &&- \I e &&\int {\rm d} \xi \xi &&\boldsymbol{B} \left( \boldsymbol{x}+\I \xi \boldsymbol{\nabla}_p,t \right) &&\times \boldsymbol{\nabla}_p. \label{eq_FVHW_Der2}  
\end{alignat} 

In order to obtain transport equations in scalar-valued quantities an expansion of the matrix-valued Wigner function \eqref{equ:WFunc} in terms of Pauli matrices is in order
\begin{equation}
\mathcal {W}^{\rm FV} \left( t, \boldsymbol{x}, \boldsymbol{p} \right) = \frac{1}{2} \left( \mathbbm{1} \mathbbm{f} + \sigma_1 \mathbbm{g} + \sigma_2 \mathbbm{h} + \sigma_3 \mathbbm{k} \right). \label{equ:wigner_sc}
\end{equation}
In this way, we obtain the transport equations for the components of the Feshbach-Villars Wigner function 
  \begin{alignat}{5}
    &m D_t \mathbbm{f} && +\left( \frac{1}{4} \boldsymbol{D}^2 - \boldsymbol{\Pi}^2 \right) \mathbbm{h} && +\boldsymbol{\Pi} \cdot \boldsymbol{D} \ \mathbbm{k} && &&= 0, \label{eq_4_1} \\
    &m D_t \mathbbm{g} && -\left( \frac{1}{4} \boldsymbol{D}^2 - \boldsymbol{\Pi}^2 \right) \mathbbm{h} && -\boldsymbol{\Pi} \cdot \boldsymbol{D} \ \mathbbm{k} && +2m^2 \mathbbm{h} &&= 0, \\    
    &m D_t \mathbbm{h} && +\left( \frac{1}{4} \boldsymbol{D}^2 - \boldsymbol{\Pi}^2 \right) \left( \mathbbm{f}+\mathbbm{g} \right) && && -2m^2 \mathbbm{g} &&= 0, \\      
    &m D_t \mathbbm{k} && && +\boldsymbol{\Pi} \cdot \boldsymbol{D} \ \left( \mathbbm{f} + \mathbbm{g} \right) \hspace{-5cm} && &&= 0. \label{eq_4_2}
  \end{alignat} 
Similarly to the Dirac-Heisenberg-Wigner formalism, we employ vacuum initial conditions \cite{FVHW}
\begin{alignat}{3}
  \mathbbm{f}_{\rm vac} \left(\boldsymbol{p} \right) = \frac{1}{2} \left( \frac{m}{\sqrt{m^2 + \boldsymbol{p}^2}} + \frac{\sqrt{m^2 + \boldsymbol{p}^2}}{m} \right), \quad \mathbbm{h}_{\rm vac}  = 0, && \\ 
  \mathbbm{g}_{\rm vac} \left(\boldsymbol{p} \right) = \frac{1}{2} \left( \frac{m}{\sqrt{m^2 + \boldsymbol{p}^2}} - \frac{\sqrt{m^2 + \boldsymbol{p}^2}}{m} \right), \quad \mathbbm{k}_{\rm vac}  = 0. &&  
\end{alignat}

\pagebreak
The particle distribution function is given by 
\begin{multline}
n^{FV} \left( \boldsymbol{x}, \boldsymbol{p} \right) = \\
\phantom{+} \frac{1}{2} \left( \frac{\sqrt{m^2 + \boldsymbol{p}^2}}{m} + \frac{m - \boldsymbol{\nabla^2}/(4m) }{\sqrt{m^2 + \boldsymbol{p}^2}} \right) \left( \mathbbm{f} - \mathbbm{f}_{\rm vac} \right) 
\\
\hspace{1.cm} + \frac{1}{2} \left( \frac{\sqrt{m^2 + \boldsymbol{p}^2}}{m} - \frac{m + \boldsymbol{\nabla^2}/(4m)}{\sqrt{m^2 + \boldsymbol{p}^2}} \right) \left( \mathbbm{g} - \mathbbm{g}_{\rm vac} \right)
 \label{equ:nF}
\end{multline}
and the particle momentum spectrum is obtained through
\begin{equation}
 n^{FV} \left( \boldsymbol{p} \right) = \int \frac{{\rm d}^3 {\boldsymbol x}}{\left( 2 \pi \right)^3} ~ n^{FV} \left( \boldsymbol{x}, \boldsymbol{p} \right).
 \label{equ:nnF}
\end{equation} 

\section{Transverse fields}
\label{sec:transverse}

Within the Heisenberg-Wigner formalism, Eqs. \eqref{eq_DHW1}-\eqref{eq_DHW2} as well as Eqs. \eqref{eq_4_1}-\eqref{eq_4_2}, the transport equations are generally formulated on a $2n$-dimensional domain where $n$ is the spatial dimension of the system. Hence, in order to solve the differential equations one is either forced to focus on lower-dimensional problems \cite{Hebenstreit, KohlfurstTech} or disregard important aspects of a fields' characteristics, e.g., it carrying linear momentum \cite{HebenstreitRelate}.

In this manuscript, we want to discuss the class of transverse field. To be more specific, we consider field configurations where all photons propagate in the same direction such that electric field $\hat{\boldsymbol{e}}_E$, magnetic field $\hat{\boldsymbol{e}}_E$ and the fields' propagation direction $\hat{\boldsymbol{e}}_k$ form the relations
\begin{equation}
\hat{\boldsymbol{e}}_E \cdot \hat{\boldsymbol{e}}_B = \hat{\boldsymbol{e}}_E \cdot \hat{\boldsymbol{e}}_k = \hat{\boldsymbol{e}}_B \cdot \hat{\boldsymbol{e}}_k = 0,\quad \hat{\boldsymbol{e}}_E \times \hat{\boldsymbol{e}}_B = \hat{\boldsymbol{e}}_k. \label{eq:Transv}
\end{equation}
These assumptions are very well valid when describing ordinary laser fields, where the transversal component of the wave vector is small compared to the parallel component ${\boldsymbol k_\perp}^2 \ll {\boldsymbol k}^2$. For example, Gaussian laser beams that are not extremely tightly focussed fall in this category.


Without loss of generality we assume photon propagation in $z$-direction; wave vector $\boldsymbol{k} = \left(0,0,k_z \right)$. Due to the relations $\boldsymbol{k} \perp \boldsymbol{E}$ and $\boldsymbol{k} \perp \boldsymbol{B}$ \eqref{eq:Transv} this further implies that photon polarization is limited to the $xy$-plane. Consequently, vector potential and fields are given by
\begin{equation}
 \boldsymbol{A} = \begin{pmatrix} A_x (t,z) \\ A_y (t,z) \\ 0 \end{pmatrix},\ \boldsymbol{E} = \begin{pmatrix} E_x (t,z) \\ E_y (t,z) \\ 0 \end{pmatrix},\ \boldsymbol{B} = \begin{pmatrix} B_x (t,z) \\ B_y (t,z) \\ 0 \end{pmatrix}, \label{equ:Field}
\end{equation}
where due to $\boldsymbol{E} = - \partial_t \boldsymbol{A}$ and $\boldsymbol{B} = \boldsymbol{\nabla}_x \times \boldsymbol{A}$ we have
\begin{align}
 & E_x = -\partial_t A_x, \quad && E_y = -\partial_t A_y, \\
 & B_x = -\partial_z A_y, \quad && B_y = +\partial_z A_x. \label{eq:EB}
\end{align}

The crucial observation is that due to the fact that the quantities in Eq. \eqref{equ:Field} do not depend on $x$ or $y$ we can make an ansatz for the spinors
\begin{equation}
 \Psi \left(t, x,y,z \right) = \ee^{\I q_x x + \I q_y y} \ \psi \left(t,z \right)
\end{equation}
in the QED Lagrangian \eqref{equ:Lag} as well as
\begin{equation}
 \Phi \left(t, x,y,z \right) = \ee^{\I q_x x + \I q_y y} \ \varphi \left(t,z \right)
\end{equation}
in case of scalar particles \eqref{equ:KG}.

Note that even if the vector potential or the fields show a dependence in $x$ or $y$, this ansatz can still be justified under the condition that the dependency is weak and can be absorbed into the overall field strength. In such a case the variables $x$ and $y$ are not treated as coordinates in physical space but as mere additional external parameters. Under such circumstances it is still possible to retain enough information to obtain an accurate prediction for the total particle yield, c.f. the dipole or locally-homogeneous approximation in Appendix \ref{App:LHA}.

Factoring out $x$- and $y$-components of the wave function has a profound impact on the Wigner operator, see Eqs. \eqref{equ:W} and \eqref{equ:Wf}. The resulting operators $\hat{\mathcal W}_{\alpha \beta}^{\perp}$ and $\hat{\mathcal W}_{\alpha \beta}^{\rm FV,\perp}$ assume the forms
\begin{multline}
 \hat{\mathcal W}_{\alpha \beta}^{\perp} \left( r_0, r_z , p \right) =
 \frac{1}{2} \int {\rm d}^4 s \ \ee^{\mathrm{i} p_0 s_0 - \mathrm{i} \boldsymbol{p} \cdot \boldsymbol{s}} \ \ee^{\I q_x s_x + \I q_y s_y} \\
 \times \ee^{\mathrm{i}e \int_{-1/2}^{1/2} {\rm d} \xi \ \left( A_0 \left(r_0+ \xi s_0, r_z+ \xi s_z \right) s_0 - \boldsymbol{A} \left(r_0+ \xi s_0, r_z+ \xi s_z \right) \cdot \boldsymbol{s} \right)} \\ 
 \times \left[ \bar {\psi}_\beta \left( r_0 - s_0/2, r_z - s_z/2 \right), \right. \\
 \left. {\psi}_\alpha \left( r_0 + s_0/2, r_z + s_z/2 \right) \right], \label{eq:W_perp_s}
\end{multline}
\begin{multline}
 \hat{\mathcal W}_{\alpha \beta}^{\rm FV, \perp} \left( t, z , p \right) = \\
 \frac{1}{2} \int {\rm d}^3 s \ \ee^{- \mathrm{i} \boldsymbol{p} \cdot \boldsymbol{s}} \ \ee^{\I q_x s_x + \I q_y s_y}
 \ \ee^{-\mathrm{i}e \int_{-1/2}^{1/2} {\rm d} \xi \ \boldsymbol{A} \left(t, z+ \xi s_z \right) \cdot \boldsymbol{s} } \\ 
 \times \left[ \bar {\varphi}_\beta \left( t, z - s_z/2 \right), {\varphi}_\alpha \left( t, z + s_z/2 \right) \right], \label{eq:W_perp_sc}
\end{multline}
where we have the relative coordinates $s = \left(s_0, s_x, s_y, s_z \right)$ and the temporal and spatial center-of-mass coordinates $(r_0,r_z)$ and $(t,z)$, respectively. The altered dependencies on relative coordinates in Eqs. \eqref{eq:W_perp_s}-\eqref{eq:W_perp_sc} makes it possible to evaluate the Fourier transforms with respect to $s_x$ and $s_y$ analytically. In this way, we find that the Wigner operators for transverse fields scale as per
\begin{multline}
 \hat{\mathcal W}_{\alpha \beta}^{\perp} \left( r_0, r_z , p \right) \propto \\
 \delta \left(p_x - q_x + e \int_{-1/2}^{1/2} {\rm d} \xi \ A_x \left(r_0 + \xi s_0, r_z + \xi s_z \right) \right)  \\
\times \delta \left(p_y - q_y + e \int_{-1/2}^{1/2} {\rm d} \xi \ A_y \left(r_0 + \xi s_0, r_z + \xi s_z \right) \right), \label{equ:Wperp}
\end{multline}
\begin{multline}
 \hat{\mathcal W}_{\alpha \beta}^{\rm FV,\perp} \left( t, z , p \right) \propto \\
 \delta \left(p_x - q_x + e \int_{-1/2}^{1/2} {\rm d} \xi \ A_x \left(t, z + \xi s_z \right) \right)  \\
\times \delta \left(p_y - q_y + e \int_{-1/2}^{1/2} {\rm d} \xi \ A_y \left(t, z + \xi s_z \right) \right). \label{equ:Wperp_sc}
\end{multline}
The Dirac delta functions basically display minimal coupling in the Wigner function approach, c.f. to lowest order
\begin{equation}
 p_x = q_x - e A_x(t,z), \quad p_y = q_y - e A_y(t,z).
\end{equation}

Equations \eqref{equ:Wperp}-\eqref{equ:Wperp_sc} perfectly demonstrate that transversal fields, when evaluated in terms of the general field notations, Eqs. \eqref{eq_DHW1}-\eqref{eq_DHW2} or Eqs. \eqref{eq_4_1}-\eqref{eq_4_2}, are bound to constraints. As such evaluation of the full transport equations is inefficient. Thus we seek for a way to incorporate the constraint equations into the transport equations. This, however, is not a trivial task as the time evolution of the Wigner components is given in phase-space, $\boldsymbol{x}$ and $\boldsymbol{p}$, while Eqs. \eqref{equ:Wperp} and \eqref{equ:Wperp_sc} suggest a relation between ordinary spatial coordinates $z$ and relative coordinates $s_z$. 


\subsection{Minimal coupling within the transport equations}
\label{sec:MinCoup}

There exist two possible pathways in order to obtain a lower-dimensional domain on which to solve the system of transport equations. Option I is to write down the Wigner operator in the form of a transverse Wigner operator as in Eq. \eqref{equ:Wperp} or Eq. \eqref{equ:Wperp_sc} and derive a system of transport equations. The alternative option is to identify the transformation that allows us to rewrite the system of transport equations, Eqs. \eqref{eq_DHW1}-\eqref{eq_DHW2} or \eqref{eq_4_1}-\eqref{eq_4_2}, and especially the differential operators, Eqs. \eqref{eqn2_1}-\eqref{eqn2_3} or \eqref{eq_FVHW_Der1}-\eqref{eq_FVHW_Der2}, into a more compact form.

In this manuscript we will pursue the second option. More specifically, we will incorporate minimal coupling for the special class of fields given by Eq. \eqref{equ:Field}. The decisive point in understanding the procedure is to realize that it is not the transport equations that have to be altered but the differential operators.
To be more specific, for fields of the form of Eq. \eqref{equ:Field} the pseudo-differential operators take on the form
\begin{align}
 & D_t && = \partial_t && + e \int {\rm d} \xi && \\
 & && \times \left( E_x \left( z+\I \xi \partial_{p_z},t \right) ~ \partial_{p_x} +  E_y \left( z+\I \xi \partial_{p_z},t \right) ~ \partial_{p_y} \right), \hspace{-10cm} && \notag \\ \label{eqn3_1}
 & D_1 && = &&+ e \int {\rm d} \xi \ && B_y \left( z+\I \xi \partial_{p_z},t \right) \partial_{p_z}, \\
 & D_2 && = &&- e \int {\rm d} \xi \ && B_x \left( z+\I \xi \partial_{p_z},t \right) \partial_{p_z}, \\   
 & D_3 && = \partial_z && + e \int {\rm d} \xi \\
 & && \times \left( B_x \left( z+\I \xi \partial_{p_z},t \right) \partial_{p_y} - B_y \left( z+\I \xi \partial_{p_z},t \right) \partial_{p_x} \right), \hspace{-10cm} \notag \\   
 & \Pi_1 && = p_x &&- \I e \int {\rm d} \xi \ \xi \ && B_y \left( z+\I \xi \partial_{p_z},t \right) \partial_{p_z}, \\ 
 & \Pi_2 && = p_y &&+ \I e \int {\rm d} \xi \ \xi \ && B_x \left( z+\I \xi \partial_{p_z},t \right) \partial_{p_z}, \\  
 & \Pi_3 && = p_z && - \I e \int {\rm d} \ \xi \ \xi && \\
 & && \times \left( B_x \left( z+\I \xi \partial_{p_z},t \right) \partial_{p_y} - B_y \left( z+\I \xi \partial_{p_z},t \right) \partial_{p_x} \right). \hspace{-10cm} && \notag \label{eqn3_2}     
\end{align}
The differential operators $\partial_{p_z}$ are only formally given in terms of arguments in the fields. Instead, terms of the form $z+\I \xi \partial_{p_z}$ should be viewed as couplings with respect to the relative coordinate. The identity transformation 
\pagebreak
\begin{multline}
 \mathcal{F}_{p_z}^{-1} \Big\{ \mathcal{F}_{p_z} \big\{ \partial_{p_z}\ f(p_z) \big\} \Big\} = \\
 \mathcal{F}_{p_z}^{-1} \Big\{ -\I s_z \ \mathcal{F}_{p_z} \big\{ f(p_z) \big\} \Big\} = \partial_{p_z}\ f(p_z),
\end{multline}
where we first perform a Fourier transform from $p_z$ to $s_z$ and then an inverse Fourier transform from $s_z$ to $p_z$, reveals this connection.

It is only at this stage, after performing the Fourier transform from $p_z$ to $s_z$ but before evaluating the inverse Fourier transform, that a coupling in spatial coordinates and relative coordinates makes sense. To be more specific, we introduce the mappings
\begin{alignat}{6}
 & p_x && = q_x - e && \int {\rm d} \xi \ && A_x \left( z + \xi s_z ,t \right), \label{eq:px} \\
 & p_y && = q_y - e && \int {\rm d} \xi \ && A_y \left( z + \xi s_z ,t \right), \label{eq:py}
\end{alignat}
coupling the transverse momenta $p_x$, $p_y$ to the vector potential and thus relative coordinates $s_z$. Establishing such a connection facilitates a switch in representation from kinetic momenta $p_x$, $p_y$ to canonical momenta $q_x$, $q_y$. As a result, applying the mapping to the differential operators in combination with a switch to canonical momenta and thus a transformation in derivatives, yields pseudo-differential operators $\cal O$ that operate under the same principle
\begin{equation}
 \mathcal{F}^{-1}_{p_z} \left\{ {\cal O} \ \mathcal{F}_{p_z} \left\{ \mathbbm{w} \right\} \right\}.
\end{equation}
In detail, the new operators are given by
\begin{alignat}{7}
 & D_t && = \partial_t, && && && \label{eqn5_1} \\   
 & D_1 && = && &&- \I e &&\int {\rm d} \xi \ && B_y \left( z+ \xi s_z ,t \right) ~ s_z , \\ 
 & D_2 && = && && +\I e &&\int {\rm d} \xi \ && B_x \left( z+ \xi s_z ,t \right) ~ s_z , \\  
 & D_3 && = \partial_z && && && \\   
 & \Pi_1 && = q_x && && && \\
 & - e \int {\rm d} \xi \ A_x \left( z+ \xi s_z ,t \right) - e\int {\rm d} \xi \ \xi \ B_y \left( z+ \xi s_z ,t \right) ~ s_z, \hspace{-8cm} && && && \notag  \\ 
 & \Pi_2 && = q_y && && && \\
 & - e \int {\rm d} \xi \ A_y \left( z+ \xi s_z ,t \right) + e \int {\rm d} \xi \ \xi \ B_x \left( z+ \xi s_z ,t \right) ~ s_z,  \hspace{-8cm} && && && \notag \\  
 & \Pi_3 && = - \I \partial_{s_z}, && && && && \label{eqn5_2} \hspace{4cm}     
\end{alignat}  
where all traces of the derivative operators $\partial_{p_x}$ and $\partial_{p_y}$ have been successfully removed. In particular, the time derivative $D_t$ as well as the derivatives in direction of propagation $D_3,~ \Pi_3$ are now given in terms of simple non-integro differential operators. 

Furthermore, due to the special form of the terms on the right-hand side we can even integrate out the dependency on the parameter $\xi$. To this end we substitute the terms $B_x$ and $B_y$ in Eqs. \eqref{eqn5_1}-\eqref{eqn5_2} by $-\partial_z A_y$ and $\partial_z A_x$, respectively. Partial integration then leads to
\pagebreak
\begin{alignat}{6}
  & && && - \I e &&\int_{-1/2}^{1/2} {\rm d} \xi \ && \partial_z A_x \left( z+ \xi s_z ,t \right) ~ s_z = \hspace{2cm} \\
  & && && && \hspace{1cm} -\I e \left\{ A_x \left( z + \frac{s_z}{2},t \right) - A_x \left( z- \frac{s_z}{2},t \right) \right\},  \hspace{-8cm} && \notag \\ 
 & && && - \I e &&\int_{-1/2}^{1/2} {\rm d} \xi \ && \partial_z A_y \left( z+ \xi s_z ,t \right) ~ s_z =  \hspace{2cm} \\
 & && && && \hspace{1cm} -\I e \left\{ A_y \left( z+ \frac{s_z}{2},t \right) - A_y \left( z- \frac{s_z}{2},t \right) \right\},  \hspace{-8cm} && \notag 
\end{alignat}
\begin{alignat}{6}
  & && - e && \int_{-1/2}^{1/2} {\rm d} \xi \ \Big( A_x \left( z+ \xi s_z ,t \right) &&+ \xi \ && \partial_z A_x \left( z+ \xi s_z ,t \right) ~ s_z \Big) \notag \\
  & && && = - \frac{e}{2} \left\{ A_x \left( z+ \frac{s_z}{2},t \right) + A_x \left( z- \frac{s_z}{2},t \right) \right\}, \hspace{-10cm} && && \\
  & && - e && \int_{-1/2}^{1/2} {\rm d} \xi \ \Big( A_y \left( z+ \xi s_z ,t \right)  &&+ \xi \ && \partial_z A_y \left( z + \xi s_z ,t \right) ~ s_z \Big) \notag \\
  & && && = - \frac{e}{2} \left\{ A_y \left( z+ \frac{s_z}{2},t \right) + A_y \left( z- \frac{s_z}{2},t \right) \right\}. \hspace{-10cm} && && 
\end{alignat}

To conclude, we have shown how to replace the integro-part $\int {\rm d}\xi$ of the differential operators by coupling terms in Fourier transformed space. In this regard, electric and magnetic fields are replaced by the vector potential. Of course, performing a Fourier transform still equates to computing an integral each time the operators are applied. However, no further numerical integration of, e.g., the electric field has to be performed. This opens up interesting possibilities in regard to an analytical evaluation of the equations.

\subsection{Minimal coupling within the DHW formalism}
\label{sec:MinCoupDHW}

In regard to the DHW formalism, the complete set of transport equations for transverse fields is given by 
\begin{alignat}{5}
    & \partial_t \mathbbm{s} && && -2 \boldsymbol{q} \cdot \boldsymbol{\mathbbm{t}}_\mathbbm{1} && -2 \boldsymbol{\Pi} \cdot \boldsymbol{\mathbbm{t}}_\mathbbm{1} &&= 0, \label{eq_5_1} \\
    & \partial_t \mathbbm{p} && && +2 \boldsymbol{q} \cdot \boldsymbol{\mathbbm{t}}_\mathbbm{2} && +2 \boldsymbol{\Pi} \cdot \boldsymbol{\mathbbm{t}}_\mathbbm{2} &&= -2m\mathbbm{a}_\mathbb{0},  \\
    & \partial_t \mathbbm{v}_\mathbb{0} &&+ \boldsymbol{D} \cdot \boldsymbol{\mathbbm{v}} && && &&= 0,  \\
    & \partial_t \mathbbm{a}_\mathbb{0} &&+ \boldsymbol{D} \cdot \boldsymbol{\mathbbm{a}} && && &&= +2m\mathbbm{p},  \\    
    & \partial_t \boldsymbol{\mathbbm{v}} &&+ \boldsymbol{D} \ \mathbbm{v}_\mathbb{0} && +2 \boldsymbol{q} \times \boldsymbol{\mathbbm{a}} && +2 \boldsymbol{\Pi} \times \boldsymbol{\mathbbm{a}} &&= -2m\boldsymbol{\mathbbm{t}}_\mathbbm{1},  \\    
    & \partial_t \boldsymbol{\mathbbm{a}} &&+ \boldsymbol{D} \ \mathbbm{a}_\mathbb{0} && +2 \boldsymbol{q} \times \boldsymbol{\mathbbm{v}} && +2 \boldsymbol{\Pi} \times \boldsymbol{\mathbbm{v}} &&= 0,  \\
    & \partial_t \boldsymbol{\mathbbm{t}}_\mathbbm{1} &&+ \boldsymbol{D} \times \boldsymbol{\mathbbm{t}}_\mathbbm{2} && +2 \boldsymbol{q} \ \mathbbm{s} && +2 \boldsymbol{\Pi} \ \mathbbm{s} &&= +2m\boldsymbol{\mathbbm{v}},  \\    
    & \partial_t \boldsymbol{\mathbbm{t}}_\mathbbm{2} &&- \boldsymbol{D} \times \boldsymbol{\mathbbm{t}}_\mathbbm{1} && -2 \boldsymbol{q} \ \mathbbm{p} && -2 \boldsymbol{\Pi} \ \mathbbm{p} &&= 0, \label{eq_5_2}  
  \end{alignat} 
where we have used vector notation for the momenta $\boldsymbol{q} = \begin{pmatrix} q_x , q_y , q_z \end{pmatrix}$. N.B.: As the vector potential in direction of $z$ vanishes, we write, for the sake of aesthetics, $p_z=q_z$. The corresponding differential operators are given by  
\begin{alignat}{6}
 & D_1 \mathbbm{w} && = \hspace{6.55cm} \label{eqn6_1} \\
 & -\I e \ \mathcal{F}^{-1}_{q_z} \left\{ \left[ A_x \left( z+ \frac{s_z}{2},t \right) - A_x \left( z- \frac{s_z}{2},t \right) \right] \mathcal{F}_{q_z} \left\{ \mathbbm{w} \right\} \right\} \hspace{-10cm} && \notag \\ \notag \\
 & D_2 \mathbbm{w} && = \\
 & -\I e \ \mathcal{F}^{-1}_{q_z} \left\{ \left[ A_y \left( z+ \frac{s_z}{2},t \right) - A_y \left( z- \frac{s_z}{2},t \right) \right] \mathcal{F}_{q_z} \left\{ \mathbbm{w} \right\} \right\} \hspace{-10cm} && \notag \\ 
 & \Pi_1 \mathbbm{w} && = \\
 &- \frac{e}{2} \ \mathcal{F}^{-1}_{q_z} \left\{ \left[ A_x \left( z+ \frac{s_z}{2},t \right) + A_x \left( z- \frac{s_z}{2},t \right) \right] \mathcal{F}_{q_z} \left\{ \mathbbm{w} \right\} \right\}  \hspace{-10cm} && \notag \\ 
 & \Pi_2 \mathbbm{w} && = \\
 &- \frac{e}{2} \ \mathcal{F}^{-1}_{q_z} \left\{ \left[ A_y \left( z+ \frac{s_z}{2},t \right) + A_y \left( z- \frac{s_z}{2},t \right) \right] \mathcal{F}_{q_z} \left\{ \mathbbm{w} \right\} \right\}  \hspace{-10cm} && \notag  
\end{alignat}  
where $\mathbbm{w} = \mathbbm{w} \left(z,\boldsymbol{q},t \right)$ is a placeholder for any of the Wigner components. Additionally, we have
\begin{equation}
D_3 = \partial_z \quad {\text{and}} \quad \Pi_3= 0.    
\label{eqn6_2}
\end{equation}

Vacuum initial conditions are given by
\begin{alignat}{7}
& \mathbbm{s}_{\rm vac} \left(\boldsymbol{q} \right) = -\frac{2m}{\sqrt{m^2 +
   \boldsymbol{q}^2}}, \qquad && 
\boldsymbol{\mathbbm{v}}_{\rm vac} \left(\boldsymbol{q} \right) = -\frac{2
   \boldsymbol{q}}{\sqrt{m^2 + \boldsymbol{q}^2}}, \notag \\    
& \mathbbm{p}_{\rm vac} = 0, \qquad 
\mathbbm{v}_\mathbb{0} {}_{\rm vac} = 0, \qquad  
\mathbbm{a}_\mathbb{0} {}_{\rm vac} = 0, \hspace{-3cm} \\
& \boldsymbol{\mathbbm{a}}_{\rm vac} = \boldsymbol{0}, \qquad  
\boldsymbol{\mathbbm{t}}_\mathbbm{1} {}_{\rm vac} = \boldsymbol{0}, \qquad  
\boldsymbol{\mathbbm{t}}_\mathbbm{2} {}_{\rm vac} = \boldsymbol{0}. \hspace{-3cm} \notag 
\end{alignat}

The particle distribution function and particle spectrum at asymptotic times ($t \to \infty$) become
\begin{equation}
n \left( z, \boldsymbol{q} \right) = \frac{m \left( \mathbbm{s}-\mathbbm{s}_{\rm vac} \right) + {\boldsymbol q} \cdot \left(
\boldsymbol{\mathbbm{v}}-\boldsymbol{\mathbbm{v}}_{\rm vac} \right)}{2\sqrt{m^2+\boldsymbol{q}^2}}
 \label{equ:n2}
\end{equation}
and
\begin{equation}
 n \left( \boldsymbol{q} \right) = \int \frac{{\rm d} z}{2 \pi} ~ n \left( z, \boldsymbol{q} \right),
 \label{equ:nn2}
\end{equation}
respectively.



\subsection{Minimal coupling within the FVHW formalism}
\label{sec:MinCoupFVHW}

In the context of sQED, a focus on transverse fields simplifies the transport equations and, more specifically, the differential operators as much as in the case of the Dirac-Heisenberg-Wigner formalism. In this regard, the full differential operators take on the form
\begin{multline}
 \left( \frac{1}{4} \boldsymbol{D}^2 - \Pi^2 \right) \mathbbm{w} = \left( \partial_z^2/4 - \boldsymbol{q}^2 \right) \mathbbm{w} \\
  + \mathcal{F}^{-1}_{q_z} \Bigg\{ \bigg( e q_x \left[ A_x (z + \frac{s_z}{2},t) + A_x (z - \frac{s_z}{2},t) \right] \bigg. \Bigg. \\
  \hspace{1.15cm} + e q_y \left[ A_y (z + \frac{s_z}{2},t) + A_y (z - \frac{s_z}{2},t) \right] \\
 \hspace{1.15cm} \Bigg. \bigg. - \frac{e^2}{2} \left[ A_x^2 (z + \frac{s_z}{2},t) + A_x^2 (z - \frac{s_z}{2},t) \right. \\
 \left. + A_y^2 (z + \frac{s_z}{2},t) + A_y^2 (z - \frac{s_z}{2},t) \right] \bigg) \mathcal{F}_{q_z} \left\{ \mathbbm{w} \right\} \Bigg\}, \label{equ:FVHW_DerA}
\end{multline}

\pagebreak
\begin{multline}
 \boldsymbol{\Pi} \cdot \boldsymbol{D} \ \mathbbm{w} = \left( q_z \ \partial_z \right) \mathbbm{w} \\
 + \mathcal{F}^{-1}_{q_z} \Bigg\{ \bigg(  -\I e q_x \left[ A_x (z + \frac{s_z}{2},t) - A_x (z - \frac{s_z}{2},t) \right] \bigg. \Bigg. \\
 \hspace{1.6cm} -\I e q_y \left[ A_y (z + \frac{s_z}{2},t) - A_y (z - \frac{s_z}{2},t) \right]  \bigg. \Bigg. \\
 \hspace{1.5cm} \Bigg. \bigg. + \frac{\I e^2}{2} \left[ A_x^2 (z + \frac{s_z}{2},t) - A_x^2 (z - \frac{s_z}{2},t) \right. \\
 \left. + A_y^2 (z + \frac{s_z}{2},t) - A_y^2 (z - \frac{s_z}{2},t) \right] \bigg) \mathcal{F}_{q_z} \left\{ \mathbbm{w} \right\} \Bigg\}, \label{equ:FVHW_DerB}
\end{multline}
with $\mathbbm{w} = \mathbbm{w} \left(t,z,\boldsymbol{q} \right)$.

The corresponding set of transport equations \eqref{eq_4_1}-\eqref{eq_4_2} is unchanged by the transformation. Nevertheless, we display equations of motion as well as initial conditions here again to have all important quantities at one location
\begin{alignat}{5}
    &m \partial_t \mathbbm{f} && +\left( \frac{1}{4} \boldsymbol{D}^2 - \boldsymbol{\Pi}^2 \right) \mathbbm{h} && +\boldsymbol{\Pi} \cdot \boldsymbol{D} \ \mathbbm{k} && &&= 0, \label{eq_FVHW1} \\
    &m \partial_t \mathbbm{g} && -\left( \frac{1}{4} \boldsymbol{D}^2 - \boldsymbol{\Pi}^2 \right) \mathbbm{h} && -\boldsymbol{\Pi} \cdot \boldsymbol{D} \ \mathbbm{k} && +2m^2 \mathbbm{h} &&= 0, \\    
    &m \partial_t \mathbbm{h} && +\left( \frac{1}{4} \boldsymbol{D}^2 - \boldsymbol{\Pi}^2 \right) \left( \mathbbm{f}+\mathbbm{g} \right) && && -2m^2 \mathbbm{g} &&= 0, \\      
    &m \partial_t \mathbbm{k} && && +\boldsymbol{\Pi} \cdot \boldsymbol{D} \ \left( \mathbbm{f} + \mathbbm{g} \right) \hspace{-5cm} && &&= 0. \label{eq_FVHW2}
  \end{alignat}
Vacuum initial conditions are given by  
\begin{alignat}{3}
  \mathbbm{f}_{\rm vac} \left(\boldsymbol{q} \right) = \frac{1}{2} \left( \frac{m}{\sqrt{m^2 + \boldsymbol{q}^2}} + \frac{\sqrt{m^2 + \boldsymbol{q}^2}}{m} \right), \quad \mathbbm{h}_{\rm vac}  = 0, && \\ 
  \mathbbm{g}_{\rm vac} \left(\boldsymbol{q} \right) = \frac{1}{2} \left( \frac{m}{\sqrt{m^2 + \boldsymbol{q}^2}} - \frac{\sqrt{m^2 + \boldsymbol{q}^2}}{m} \right), \quad \mathbbm{k}_{\rm vac}  = 0, &&  
\end{alignat}
and the particle distribution function takes on the form
\begin{multline}
n^{FV} \left( \boldsymbol{x}, \boldsymbol{q} \right) = \\
\phantom{+} \frac{1}{2} \left( \frac{\sqrt{m^2 + \boldsymbol{q}^2}}{m} + \frac{m - \partial_z^2/(4m)}{\sqrt{m^2 + \boldsymbol{q}^2}}  \right) \left( \mathbbm{f} - \mathbbm{f}_{\rm vac} \right) 
\\
\hspace{1.cm} + \frac{1}{2} \left( \frac{\sqrt{m^2 + \boldsymbol{q}^2}}{m} - \frac{m+\partial_z^2/(4m)}{\sqrt{m^2 + \boldsymbol{q}^2}} \right) \left( \mathbbm{g} - \mathbbm{g}_{\rm vac} \right).
\end{multline}

\section{Application: \ Bi-frequent fields within a periodic envelope}
\label{sec:Appl}


One possible application of the Heisenberg-Wigner formalism for transverse fields is given by the study of particle production rates in colliding, high-intensity waves. The exemplary field configuration we will discuss in this manuscript is given by
\begin{align}
 \begin{split}
 A_x (z \pm \frac{s_z}{2},t) &= A_{Z,x} \left( z \pm \frac{s_z}{2} \right) \times \\
 & \left( A_{1,x} \left( t \right) \ \cos \left( \omega_1 t + k_1 \left( z \pm \frac{s_z}{2} \right) \right) \right. \\
 & \left. + A_{2,x} \left( t \right) \ \cos \left( \omega_2 t - k_2 \left( z \pm \frac{s_z}{2} \right) \right) \right), \label{eq:ATZxPer}
 \end{split} \\
 \begin{split}
 A_y (z \pm \frac{s_z}{2},t) &= A_{Z,y} \left( z \pm \frac{s_z}{2} \right) \times \\
  & \left( A_{1,y} \left( t \right) \ \sin \left( \omega_1 t + k_1 \left( z \pm \frac{s_z}{2} \right) \right) \right. \\
  & \left. + A_{2,y} \left( t \right) \ \sin \left( \omega_2 t - k_2 \left( z \pm \frac{s_z}{2} \right) \right) \right), \label{eq:ATZyPer}
 \end{split} 
\end{align}
with the the spatial envelope functions $A_{Z,x} \left( z \right)$ and $A_{Z,y} \left( z \right)$, the temporal envelope functions $A_{1,x} \left( t \right)$, $A_{1,y} \left( t \right)$, $A_{2,x} \left( t \right)$ and $A_{2,y} \left( t \right)$ as well as photon energies $\omega_1$, $\omega_2$ and photon momenta $k_1$, $k_2$.

Note that we use spatial envelope functions $A_{Z,x} \left( z \right)$ and $A_{Z,y} \left( z \right)$, because we want to discuss field configurations where the interaction region of, e.g., two colliding pulses is such that the spatial finiteness cannot be ignored. We further assume, that an expansion of the spatial envelope function in terms of $z$ would only give a crude approximation and is therefore not applicable.

The impact of the envelope is not to be underestimated as Refs. \cite{HeinzlIlderton,KingEnv} have shown. Nevertheless, as our goal at this point is to demonstrate the power of incorporating minimal coupling at the operator level, we focus on field configurations that exhibit a periodic envelope function. The reason is that Fourier transforming a sine or cosine function yields a delta distribution as a result. Hence, spatially oscillating field profiles are especially well suited to be incorporated into the Heisenberg-Wigner formalism, see Sec.\ref{sec:Outlook} for an outlook on the possibilities of modeling pulse shapes including a discussion on ways to implement non-periodic envelope functions.  




In this context, we assume an envelope function of the form of
\begin{multline}
 A_{Z,x} \left( z \pm \frac{s_z}{2} \right) = A_{Z,y} \left( z \pm \frac{s_z}{2} \right) = \\
 \cos \left( \frac{1}{\lambda} \left( z \pm \frac{s_z}{2} \right) \right)^2. \label{eq:ATZEnv}
\end{multline}
We have not specified any constraints on the width parameter $\lambda$. However, if clear periodicity of the field configuration is to be obtained,
the subcycle oscillation period has to match the period of the envelope function. In this regard, it should also be clarified that any even exponent is equally well suited for the task of containing the characteristics of the field. The larger the exponent the better the envelope function approximates a flat top pulse and thus the more impact it has on the momentum transfer from the field to particles.

If a particle number is to be extracted, a proper normalization has to be done after evaluating the transport equations \cite{KohlfurstDirac}.

Additionally, the field configuration \eqref{eq:ATZxPer}-\eqref{eq:ATZyPer} is formulated in a very general way. Hence, a variety of simpler configurations can be derived from the final set of equations by choosing suitable amplitudes and frequencies. For example, in the limit 
\begin{align}
& A_{1,x} \left( t \right) = A_{2,x} \left( t \right) = A_{x} \left( t \right),\\  
& A_{1,y} \left( t \right) = A_{2,y} \left( t \right) = A_{y} \left( t \right), \\
& \omega_1 = \omega_2 = \omega,~ k_1 = k_2 = k,~ \lambda \to \infty,
\end{align}
the coupling operators describing a standing wave configuration are recovered. Such, and other applications are given in the appendix, see Appendix \ref{App:LHA} for locally homogeneous fields, Appendix \ref{App:Assist} for assisting potentials, Appendix \ref{App:Stand} for a standing wave pattern and Appendix \ref{App:Bi} for the interaction of bi-frequent fields.

\subsection{Dirac-Heisenberg-Wigner formalism}
\label{sec:Appl_DHW}

Within the DHW formalism for transverse fields, employing a field shape of the form of Eqs. \eqref{eq:ATZxPer}-\eqref{eq:ATZyPer} with envelope Eq. \eqref{eq:ATZEnv}, turns the modification factors in Eqs. \eqref{eqn6_1}-\eqref{eqn6_2} into 
\begin{widetext}
\begin{align}
 \begin{split}
 \left[ A_x \right. & \left. \left( z+ \frac{s_z}{2},t \right) - A_x \left( z- \frac{s_z}{2},t \right) \right] =  \label{eq:A1A21e} \\
 & +\cos \left( \frac{s_z + 2z}{2 \lambda} \right)^2 \ \left( A_{1,x} (t) \ \cos \left( \omega_1 t + k_1 z + \frac{k_1 s_z}{2} \right) + A_{2,x} (t) \ \cos \left( \omega_2 t - k_2 z - \frac{k_2 s_z}{2} \right) \right) \\
 & -\cos \left( \frac{s_z - 2z}{2 \lambda} \right)^2 \ \left( A_{1,x} (t) \ \cos \left( \omega_1 t + k_1 z - \frac{k_1 s_z}{2} \right) + A_{2,x} (t) \ \cos \left( \omega_2 t - k_2 z + \frac{k_2 s_z}{2} \right) \right), 
\end{split} \\
\begin{split}
 \left[ A_y \right. & \left. \left( z+ \frac{s_z}{2},t \right) - A_y \left( z- \frac{s_z}{2},t \right) \right] = \label{eq:A1A22e} \\
 & +\cos \left( \frac{s_z + 2z}{2 \lambda} \right)^2 \ \left( A_{1,x} (t) \ \sin \left( \omega_1 t + k_1 z + \frac{k_1 s_z}{2} \right) + A_{2,x} (t) \ \sin \left( \omega_2 t - k_2 z - \frac{k_2 s_z}{2} \right) \right) \\
 & -\cos \left( \frac{s_z - 2z}{2 \lambda} \right)^2 \ \left( A_{1,x} (t) \ \sin \left( \omega_1 t + k_1 z - \frac{k_1 s_z}{2} \right) + A_{2,x} (t) \ \sin \left( \omega_2 t - k_2 z + \frac{k_2 s_z}{2} \right) \right), 
\end{split} \\
 \begin{split}
 \left[ A_x \right. & \left. \left( z+ \frac{s_z}{2},t \right) + A_x \left( z- \frac{s_z}{2},t \right) \right] = \label{eq:A1A23e} \\
 & +\cos \left( \frac{s_z + 2z}{2 \lambda} \right)^2 \ \left( A_{1,y} (t) \ \cos \left( \omega_1 t + k_1 z + \frac{k_1 s_z}{2} \right) + A_{2,y} (t) \ \cos \left( \omega_2 t - k_2 z - \frac{k_2 s_z}{2} \right) \right) \\
 & +\cos \left( \frac{s_z - 2z}{2 \lambda} \right)^2 \ \left( A_{1,y} (t) \ \cos \left( \omega_1 t + k_1 z - \frac{k_1 s_z}{2} \right) + A_{2,y} (t) \ \cos \left( \omega_2 t - k_2 z + \frac{k_2 s_z}{2} \right) \right), 
 \end{split} \\
 \begin{split}
 \left[ A_y \right. & \left. \left( z+ \frac{s_z}{2},t \right) + A_y \left( z- \frac{s_z}{2},t \right) \right] = \label{eq:A1A24e} \\
 & +\cos \left( \frac{s_z + 2z}{2 \lambda} \right)^2 \ \left( A_{1,y} (t) \ \sin \left( \omega_1 t + k_1 z + \frac{k_1 s_z}{2} \right) + A_{2,y} (t) \ \sin \left( \omega_2 t - k_2 z - \frac{k_2 s_z}{2} \right) \right) \\
 & +\cos \left( \frac{s_z - 2z}{2 \lambda} \right)^2 \ \left( A_{1,y} (t) \ \sin \left( \omega_1 t + k_1 z - \frac{k_1 s_z}{2} \right) + A_{2,y} (t) \ \sin \left( \omega_2 t - k_2 z + \frac{k_2 s_z}{2} \right) \right). 
 \end{split} 
\end{align}

Consequently, Fourier transforms can be performed analytically, thus we obtain a discrete coupling mechanism 
\begin{alignat}{6}
 & D_1 \mathbbm{w} && = \sum_{n=-1}^1 \binom{2}{n+1} \times && && && \label{eqn12b_1} \\
 & && && \hspace{-0.5 cm} \Bigg\{ \Bigg. -\frac{e A_{1,x} \left( t \right) }{4} \sin \left(\omega_1 t + k_1 z + \frac{2nz}{\lambda} \right) \ \left[ \mathbbm{w} \left( z, q_z+\frac{k_1}{2} +\frac{n}{\lambda} \right) - \mathbbm{w} \left( z, q_z - \frac{k_1}{2} - \frac{n}{\lambda} \right) \right] && && \notag \\
 & && && \hspace{-0.15 cm} +\frac{e A_{2,x} \left( t \right) }{4} \sin \left(\omega_2 t - k_2 z + \frac{2nz}{\lambda} \right) \ \left[ \mathbbm{w} \left( z, q_z+\frac{k_2}{2} - \frac{n}{\lambda} \right) - \mathbbm{w} \left( z, q_z - \frac{k_2}{2} + \frac{n}{\lambda} \right) \right] \Bigg. \Bigg\}, && && \notag \\ \notag
 \end{alignat}
\begin{alignat}{6} 
 & D_2 \mathbbm{w} && = \sum_{n=-1}^1 \binom{2}{n+1} \times && && && \label{eqn12b_3} \\
 & && && \hspace{-0.5 cm} \Bigg\{ \Bigg. +\frac{e A_{1,y} \left( t \right) }{4} \cos \left(\omega_1 t + k_1 z + \frac{2nz}{\lambda} \right) \ \left[ \mathbbm{w} \left( z, q_z+\frac{k_1}{2} +\frac{n}{\lambda} \right) - \mathbbm{w} \left( z, q_z - \frac{k_1}{2} - \frac{n}{\lambda} \right) \right] && \notag \\
 & && && \hspace{-0.15 cm} -\frac{e A_{2,y} \left( t \right) }{4} \cos \left(\omega_2 t - k_2 z + \frac{2nz}{\lambda} \right) \ \left[ \mathbbm{w} \left( z, q_z+\frac{k_2}{2} - \frac{n}{\lambda} \right) - \mathbbm{w} \left( z, q_z - \frac{k_2}{2} + \frac{n}{\lambda} \right) \right] \Bigg. \Bigg\}, && && \notag \\  
 & \Pi_1 \mathbbm{w} && = \sum_{n=-1}^1 \binom{2}{n+1} \times && && && \label{eqn12b_4} \\
 & && && \hspace{-0.5 cm} \Bigg\{ \Bigg. -\frac{e A_{1,x} \left( t \right) }{8} \cos \left(\omega_1 t + k_1 z + \frac{2nz}{\lambda} \right) \ \left[ \mathbbm{w} \left( z, q_z+\frac{k_1}{2} +\frac{n}{\lambda} \right) + \mathbbm{w} \left( z, q_z - \frac{k_1}{2} - \frac{n}{\lambda} \right) \right] && && \notag \\
 & && && \hspace{-0.15 cm} -\frac{e A_{2,x} \left( t \right) }{8} \cos \left(\omega_2 t - k_2 z + \frac{2nz}{\lambda} \right) \ \left[ \mathbbm{w} \left( z, q_z+\frac{k_2}{2} - \frac{n}{\lambda} \right) + \mathbbm{w} \left( z, q_z - \frac{k_2}{2} + \frac{n}{\lambda} \right) \right] \Bigg. \Bigg\}, && && \notag \\  
 & \Pi_2 \mathbbm{w} && = \sum_{n=-1}^1 \binom{2}{n+1} \times && && && \label{eqn12b_5} \\
 & && && \hspace{-0.5 cm} \Bigg\{ \Bigg. -\frac{e A_{1,y} \left( t \right) }{8} \sin \left(\omega_1 t + k_1 z + \frac{2nz}{\lambda} \right) \ \left[ \mathbbm{w} \left( z, q_z+\frac{k_1}{2} +\frac{n}{\lambda} \right) + \mathbbm{w} \left( z, q_z - \frac{k_1}{2} - \frac{n}{\lambda} \right) \right] && && \notag \\
 & && && \hspace{-0.15 cm} -\frac{e A_{2,y} \left( t \right) }{8} \sin \left(\omega_2 t - k_2 z + \frac{2nz}{\lambda} \right) \ \left[ \mathbbm{w} \left( z, q_z+\frac{k_2}{2} - \frac{n}{\lambda} \right) + \mathbbm{w} \left( z, q_z - \frac{k_2}{2} + \frac{n}{\lambda} \right) \right] \Bigg. \Bigg\}. && && \notag 
\end{alignat}  
\end{widetext}
Furthermore, we have $D_3=\partial_z$ and a vanishing operator $\Pi_3=0$.

We observe that there are two factors that weight in in these discrete coupling operators. First, there is a momentum transfer coming from the plane-wave features, see the terms $\left( k_1 s_z \right)/2$ and $\left( k_2 s_z \right)/2$ in Eqs. \eqref{eq:A1A21e}-\eqref{eq:A1A24e} that translate into momentum couplings $q_z \pm k_1/2$ and $q_z \pm k_2/2$, respectively. The second contribution is due to the envelope function which adds an additional layer of complexity scaling as per $\propto 1/\lambda$. 

At this level, the differential operators can be readily applied to the transport equations having the advantage that no numerical evaluation of Fourier transforms have to be performed at all. Conceptually, such a formulation of the Heisenberg-Wigner formalism is exceptionally intriguing as calculations can be performed entirely in coordinate-momentum-space, see Sec. \ref{sec:MinCoupDHW} for the transport equations, vacuum initial conditions as well as a definition of observables.

Nevertheless, to put the possibilities of the Heisenberg-Wigner formalism for transverse fields to full display, we further improve on the representation of the transport equations. Identifying that in Eqs. \eqref{eqn12b_1}-\eqref{eqn12b_5} the dependency of the spatial coordinate $z$ is entirely expressed in terms of sine and cosine functions, performing yet another Fourier transform we transfer the system to a complete energy-momentum-based formalism. In this regard, $\boldsymbol{q}$ denotes momenta in phase-space and instead of the spatial coordinate $z$ we have the variable $K$ denoting energy-momentum channels. To be more specific, assuming that the period of envelope function and subcycle oscillations match we can expand the Wigner components in terms of a Fourier series.


This yields the coupling terms
\begin{widetext}
\begin{alignat}{6}
 & && D_1 \mathbbm{w} = && \sum_{n=-1}^1 \ \binom{2}{n+1} && \ \frac{\I e A_{1,x} \left( t \right) }{8} \times && && \label{eqn13_1} \\
 & && && && \left( + \exp \left(+\I \omega_1 t \right) \ \left[ \tilde{\mathbbm{w}} \left( K - k_1 - \frac{2n}{\lambda}, q_z+\frac{k_1}{2} + \frac{n}{\lambda} \right) - \tilde{\mathbbm{w}} \left( K -k_1 - \frac{2n}{\lambda} , q_z - \frac{k_1}{2} - \frac{n}{\lambda} \right) \right] \right. && && \notag \\
 & && && && \hspace{0.3cm} \left. - \exp \left(-\I \omega_1 t \right) \ \left[ \tilde{\mathbbm{w}} \left( K + k_1 + \frac{2n}{\lambda}, q_z+\frac{k_1}{2} +\frac{n}{\lambda} \right) - \tilde{\mathbbm{w}} \left( K + k_1 + \frac{2n}{\lambda} , q_z - \frac{k_1}{2} - \frac{n}{\lambda} \right) \right] \right) && && \notag \\ 
 & && && - \sum_{n=-1}^1 \ \binom{2}{n+1} \ \frac{\I e A_{2,x} \left( t \right) }{8} \times \hspace{-5cm} && && && \notag \\ 
 & && && && \left( + \exp \left(+\I \omega_2 t \right) \ \left[ \tilde{\mathbbm{w}} \left( K + k_2 - \frac{2n}{\lambda} , q_z + \frac{k_2}{2} - \frac{n}{\lambda} \right) - \tilde{\mathbbm{w}} \left( K + k_2 - \frac{2n}{\lambda}, q_z-\frac{k_2}{2} + \frac{n}{\lambda} \right) \right] \right. && && \notag \\
 & && && && \hspace{0.3cm} \left. - \exp \left(-\I \omega_2 t \right) \ \left[ \tilde{\mathbbm{w}} \left( K - k_2 + \frac{2n}{\lambda} , q_z + \frac{k_2}{2} - \frac{n}{\lambda} \right) - \tilde{\mathbbm{w}} \left( K - k_2 + \frac{2n}{\lambda}, q_z-\frac{k_2}{2} + \frac{n}{\lambda} \right) \right] \right), && && \notag \\ 
 & && D_2 \mathbbm{w} = && \sum_{n=-1}^1 \ \binom{2}{n+1} && \ \frac{e A_{1,y} \left( t \right) }{8} \times && && \label{eqn13_2} \\
 & && && && \left( + \exp \left(+\I \omega_1 t \right) \ \left[ \tilde{\mathbbm{w}} \left( K - k_1 - \frac{2n}{\lambda}, q_z+\frac{k_1}{2} +\frac{n}{\lambda} \right) - \tilde{\mathbbm{w}} \left( K -k_1 - \frac{2n}{\lambda} , q_z - \frac{k_1}{2} - \frac{n}{\lambda} \right) \right] \right. && && \notag \\
 & && && && \hspace{0.3cm} \left. + \exp \left(-\I \omega_1 t \right) \ \left[ \tilde{\mathbbm{w}} \left( K + k_1 + \frac{2n}{\lambda}, q_z+\frac{k_1}{2} +\frac{n}{\lambda} \right) - \tilde{\mathbbm{w}} \left( K + k_1 + \frac{2n}{\lambda} , q_z - \frac{k_1}{2} - \frac{n}{\lambda} \right) \right] \right) && && \notag \\ 
 & && && - \sum_{n=-1}^1 \ \binom{2}{n+1} \ \frac{e A_{2,y} \left( t \right) }{8} \times \hspace{-5cm} && && && \notag \\ 
 & && && && \left( + \exp \left(+\I \omega_2 t \right) \ \left[ \tilde{\mathbbm{w}} \left( K + k_2 - \frac{2n}{\lambda} , q_z + \frac{k_2}{2} - \frac{n}{\lambda} \right) - \tilde{\mathbbm{w}} \left( K + k_2 - \frac{2n}{\lambda}, q_z-\frac{k_2}{2} + \frac{n}{\lambda} \right) \right] \right. && && \notag \\
 & && && && \hspace{0.3cm} \left. + \exp \left(-\I \omega_2 t \right) \ \left[ \tilde{\mathbbm{w}} \left( K - k_2 + \frac{2n}{\lambda} , q_z + \frac{k_2}{2} - \frac{n}{\lambda} \right) - \tilde{\mathbbm{w}} \left( K - k_2 + \frac{2n}{\lambda}, q_z-\frac{k_2}{2} + \frac{n}{\lambda} \right) \right] \right), && && \notag  
 \end{alignat}
 \begin{alignat}{6}
 & && \Pi_1 \mathbbm{w} = && -\sum_{n=-1}^1 \ \binom{2}{n+1} && \ \frac{e A_{1,x} \left( t \right) }{16} \times && && \label{eqn13_4} \\
 & && && && \left( + \exp \left(+\I \omega_1 t \right) \ \left[ \tilde{\mathbbm{w}} \left( K - k_1 - \frac{2n}{\lambda}, q_z+\frac{k_1}{2} +\frac{n}{\lambda} \right) + \tilde{\mathbbm{w}} \left( K -k_1 - \frac{2n}{\lambda} , q_z - \frac{k_1}{2} - \frac{n}{\lambda} \right) \right] \right. && && \notag \\
 & && && && \hspace{0.3cm} \left. + \exp \left(-\I \omega_1 t \right) \ \left[ \tilde{\mathbbm{w}} \left( K + k_1 + \frac{2n}{\lambda}, q_z+\frac{k_1}{2} +\frac{n}{\lambda} \right) + \tilde{\mathbbm{w}} \left( K + k_1 + \frac{2n}{\lambda} , q_z - \frac{k_1}{2} - \frac{n}{\lambda} \right) \right] \right) && && \notag \\ 
 & && && - \sum_{n=-1}^1 \ \binom{2}{n+1} \ \frac{e A_{2,x} \left( t \right) }{16} \times \hspace{-5cm} && && && \notag \\ 
 & && && && \left( + \exp \left(+\I \omega_2 t \right) \ \left[ \tilde{\mathbbm{w}} \left( K + k_2 - \frac{2n}{\lambda} , q_z + \frac{k_2}{2} - \frac{n}{\lambda} \right) + \tilde{\mathbbm{w}} \left( K + k_2 - \frac{2n}{\lambda}, q_z-\frac{k_2}{2} + \frac{n}{\lambda} \right) \right] \right. && && \notag \\
 & && && && \hspace{0.3cm} \left. + \exp \left(-\I \omega_2 t \right) \ \left[ \tilde{\mathbbm{w}} \left( K - k_2 + \frac{2n}{\lambda} , q_z + \frac{k_2}{2} - \frac{n}{\lambda} \right) + \tilde{\mathbbm{w}} \left( K - k_2 + \frac{2n}{\lambda}, q_z-\frac{k_2}{2} + \frac{n}{\lambda} \right) \right] \right), && && \notag \\ 
 & && \Pi_2 \mathbbm{w} = && \sum_{n=-1}^1 \ \binom{2}{n+1} && \ \frac{\I e A_{1,y} \left( t \right) }{16} \times && && \label{eqn13_5} \\
 & && && && \left( + \exp \left(+\I \omega_1 t \right) \ \left[ \tilde{\mathbbm{w}} \left( K - k_1 - \frac{2n}{\lambda}, q_z+\frac{k_1}{2} +\frac{n}{\lambda} \right) + \tilde{\mathbbm{w}} \left( K -k_1 - \frac{2n}{\lambda} , q_z - \frac{k_1}{2} - \frac{n}{\lambda} \right) \right] \right. && && \notag \\
 & && && && \hspace{0.3cm} \left. - \exp \left(-\I \omega_1 t \right) \ \left[ \tilde{\mathbbm{w}} \left( K + k_1 + \frac{2n}{\lambda}, q_z+\frac{k_1}{2} +\frac{n}{\lambda} \right) + \tilde{\mathbbm{w}} \left( K + k_1 + \frac{2n}{\lambda} , q_z - \frac{k_1}{2} - \frac{n}{\lambda} \right) \right] \right) && && \notag \\ 
 & && && + \sum_{n=-1}^1 \ \binom{2}{n+1} \ \frac{\I e A_{2,y} \left( t \right) }{16} \times \hspace{-5cm} && && && \notag \\ 
 & && && && \left( + \exp \left(+\I \omega_2 t \right) \ \left[ \tilde{\mathbbm{w}} \left( K + k_2 - \frac{2n}{\lambda} , q_z + \frac{k_2}{2} - \frac{n}{\lambda} \right) + \tilde{\mathbbm{w}} \left( K + k_2 - \frac{2n}{\lambda}, q_z-\frac{k_2}{2} + \frac{n}{\lambda} \right) \right] \right. && && \notag \\
 & && && && \hspace{0.3cm} \left. - \exp \left(-\I \omega_2 t \right) \ \left[ \tilde{\mathbbm{w}} \left( K - k_2 + \frac{2n}{\lambda} , q_z + \frac{k_2}{2} - \frac{n}{\lambda} \right) + \tilde{\mathbbm{w}} \left( K - k_2 + \frac{2n}{\lambda}, q_z-\frac{k_2}{2} + \frac{n}{\lambda} \right) \right] \right). && && \notag 
\end{alignat}
\end{widetext}
Additionally, we have $D_3 \mathbbm{w} = i K \tilde{\mathbbm{w}} \left( K, q_z \right)$ and $\Pi_3 = 0$. \\
 
The transport equations then take on the form
\begin{alignat}{5}
    & \partial_t \tilde{\mathbbm{s}} && && -2 \boldsymbol{q} \cdot \tilde{\mathbbm{t}}_\mathbbm{1} && -2 \boldsymbol{\Pi} \cdot \boldsymbol{\mathbbm{t}}_\mathbbm{1} &&= 0, \label{eq_7_1} \\
    & \partial_t \tilde{\mathbbm{p}} && && +2 \boldsymbol{q} \cdot \tilde{\mathbbm{t}}_\mathbbm{2} && +2 \boldsymbol{\Pi} \cdot \boldsymbol{\mathbbm{t}}_\mathbbm{2} &&= -2m \tilde{\mathbbm{a}}_\mathbb{0},  \\
    & \partial_t \tilde{\mathbbm{v}}_\mathbb{0} &&+ \boldsymbol{D} \cdot \boldsymbol{\mathbbm{v}} && && &&= 0,  \\
    & \partial_t \tilde{\mathbbm{a}}_\mathbb{0} &&+ \boldsymbol{D} \cdot \boldsymbol{\mathbbm{a}} && && &&= +2m \tilde{\mathbbm{p}},  \\    
    & \partial_t \tilde{\boldsymbol{\mathbbm{v}}} &&+ \boldsymbol{D} \ \mathbbm{v}_\mathbb{0} && +2 \boldsymbol{q} \times \tilde{\boldsymbol{\mathbbm{a}}} && +2 \boldsymbol{\Pi} \times \boldsymbol{\mathbbm{a}} &&= -2m \tilde{\mathbbm{t}}_\mathbbm{1},  \\    
    & \partial_t \tilde{\boldsymbol{\mathbbm{a}}} &&+ \boldsymbol{D} \ \mathbbm{a}_\mathbb{0} && +2 \boldsymbol{q} \times \tilde{\boldsymbol{\mathbbm{v}}} && +2 \boldsymbol{\Pi} \times \boldsymbol{\mathbbm{v}} &&= 0,  \\
    & \partial_t \tilde{\mathbbm{t}}_\mathbbm{1} &&+ \boldsymbol{D} \times \boldsymbol{\mathbbm{t}}_\mathbbm{2} && +2 \boldsymbol{q} \ \tilde{\mathbbm{s}} && +2 \boldsymbol{\Pi} \ \mathbbm{s} &&= +2m \tilde{\boldsymbol{\mathbbm{v}}},  \\    
    & \partial_t \tilde{\mathbbm{t}}_\mathbbm{2} &&- \boldsymbol{D} \times \boldsymbol{\mathbbm{t}}_\mathbbm{1} && -2 \boldsymbol{q} \ \tilde{\mathbbm{p}} && -2 \boldsymbol{\Pi} \ \mathbbm{p} &&= 0, \label{eq_7_2}  
\end{alignat} 
where Wigner components $\mathbbm{w}$ are treated under the procedure displayed in Eqs. \eqref{eqn13_1}-\eqref{eqn13_5}. Components that are denoted as $\tilde{\mathbbm{w}} = \tilde{\mathbbm{w}} \left(K, q_z \right)$ have already been transformed.
  
In this representation, vacuum initial conditions are given by
\begin{alignat}{8}
& \tilde{\mathbbm{s}}_{\rm vac} \left(K, \boldsymbol{q} \right) && = -\frac{2m}{\sqrt{m^2 + \boldsymbol{q}^2}} \ \delta \left( K \right), \quad && 
\tilde{\mathbbm{p}}_{\rm vac} && = 0, && \\
& \tilde{\mathbbm{v}}_\mathbb{0} {}_{\rm vac} && = 0, \quad && 
\tilde{\mathbbm{a}}_\mathbb{0} {}_{\rm vac} && = 0, && \\
& \tilde{\boldsymbol{\mathbbm{v}}}_{\rm vac} \left(K, \boldsymbol{q} \right) && = -\frac{2
   \boldsymbol{q}}{\sqrt{m^2 + \boldsymbol{q}^2}} \ \delta \left( K \right), \quad && 
 \tilde{\boldsymbol{\mathbbm{a}}}_{\rm vac} && = \boldsymbol{0}, && \\ 
& \tilde{\boldsymbol{\mathbbm{t}}}_\mathbbm{1} {}_{\rm vac} && = \boldsymbol{0}, \quad && 
\tilde{\boldsymbol{\mathbbm{t}}}_\mathbbm{2} {}_{\rm vac} && = \boldsymbol{0}, \quad && 
\end{alignat}
thus only zero-modes in $K$ do not vanish at $t \to - \infty$. 
Accordingly, the particle distribution function reads
\begin{equation}
n \left( K, \boldsymbol{q} \right) = \frac{m \left( \tilde{\mathbbm{s}}-\tilde{\mathbbm{s}}_{\rm vac} \right) + {\boldsymbol q} \cdot \left( \tilde{\boldsymbol{\mathbbm{v}}} - \tilde{\boldsymbol{\mathbbm{v}}}_{\rm vac} \right)}{2\sqrt{m^2+\boldsymbol{q}^2}}. 
\end{equation}

    
\subsection{Feshbach-Villars-Heisenberg-Wigner formalism}
\label{sec:Appl_FVHW}

We have already seen how to adopt a complex field configuration into this new variation of the Heisenberg-Wigner formalism in Sec. \ref{sec:Appl_DHW} for Dirac particles. While the same mechanism applies for the FVHW formalism, the differential operators are much more involved thus obtaining the characteristic coupling terms in the Wigner components is more complicated.

Given a configuration of the form of Eqs. \eqref{eq:ATZxPer}-\eqref{eq:ATZyPer} the differential operators \eqref{equ:FVHW_DerA}-\eqref{equ:FVHW_DerB} are converted into coupling operators of the form of \cite{footnote3}  
\begin{widetext}
\begin{align}
 & \left( \frac{1}{4} \boldsymbol{D}^2 - \Pi^2 \right) \mathbbm{w} =  \label{eq:FV_Op1} \\
 & \hspace{1cm} \left( \partial_z^2/4 - \boldsymbol{q}^2 \right) \times \left[ \mathbbm{w} \left( z, q_z \right) \right] \notag \\
 \begin{split}
 & \hspace{1cm} +\sum_{i=1}^2 \ \sum_{n=-1}^1 \binom{2}{n+1} \ \frac{e}{4} \left( A_{i,x} \left( t \right) q_x \cos \left(\omega_i t + \chi_i k_i z + \frac{2nz}{\lambda} \right) \right. \\
 & \hspace{6cm} \left. + A_{i,y} \left( t \right) q_y \sin \left(\omega_i t + \chi_i k_i z + \frac{2nz}{\lambda} \right) \right) \times  \\
 & \hspace{7cm} \left[ \mathbbm{w} \left( z, q_z+\frac{\chi_i k_i}{2} +\frac{n}{\lambda} \right) + \mathbbm{w} \left( z, q_z - \frac{\chi_i k_i}{2} - \frac{n}{\lambda} \right) \right]  \notag 
 \end{split} \\
 \begin{split}
 & \hspace{1cm} -\sum_{i=1}^2 \ \sum_{j=1}^2 \ \sum_{\rho=-1,+1} \ \sum_{n=-2}^2 \binom{4}{n+2} \ \frac{e^2}{64} \Big( A_{i,x} \left( t \right) A_{j,x} \left( t \right) - \rho A_{i,y} \left( t \right) A_{j,y} \left( t \right) \Big) \times \\
 & \hspace{2cm} \cos \Big( \left(\omega_i + \rho \omega_j \right) t + \left(\chi_i k_i + \rho \chi_j k_j \right) z + \frac{2nz}{\lambda} \Big) \times  \\
 & \hspace{3.8cm} \left[ \mathbbm{w} \left( z, q_z+\frac{\chi_i k_i}{2} + \frac{\rho \chi_j k_j}{2} +\frac{n}{\lambda} \right) + \mathbbm{w} \left( z, q_z - \frac{\chi_i k_i}{2} - \frac{\rho \chi_j k_j}{2} - \frac{n}{\lambda} \right) \right],  \notag  
 \end{split} 
\end{align}
%
\begin{align}
 & \left( \boldsymbol{\Pi} \cdot \boldsymbol{D} \right) \mathbbm{w} =  \label{eq:FV_Op2} \\
 & \hspace{1cm} \left( q_z \ \partial_z \right) \times \left[ \mathbbm{w} \left( z, q_z \right) \right] \notag \\
 \begin{split}
 & \hspace{1cm} -\sum_{i=1}^2 \ \sum_{n=-1}^1 \binom{2}{n+1} \ \frac{e}{4} \left( A_{i,x} \left( t \right) q_x \sin \left(\omega_i t + \chi_i k_i z + \frac{2nz}{\lambda} \right) \right. \\
 & \hspace{6cm} \left. - A_{i,y} \left( t \right) q_y \cos \left(\omega_i t + \chi_i k_i z + \frac{2nz}{\lambda} \right) \right) \times  \\
 & \hspace{7cm} \left[ \mathbbm{w} \left( z, q_z+\frac{\chi_i k_i}{2} +\frac{n}{\lambda} \right) - \mathbbm{w} \left( z, q_z - \frac{\chi_i k_i}{2} - \frac{n}{\lambda} \right) \right]  \notag 
 \end{split} \\
 \begin{split}
 & \hspace{1cm} +\sum_{i=1}^2 \ \sum_{j=1}^2 \ \sum_{\rho=-1,+1} \ \sum_{n=-2}^2 \binom{4}{n+2} \ \frac{e^2}{64} \Big( A_{i,x} \left( t \right) A_{j,x} \left( t \right) - \rho A_{i,y} \left( t \right) A_{j,y} \left( t \right) \Big) \times \\
 & \hspace{2cm} \sin \Big( \left(\omega_i + \rho \omega_j \right) t + \left(\chi_i k_i + \rho \chi_j k_j \right) z + \frac{2nz}{\lambda} \Big) \times  \\
 & \hspace{3.8cm} \left[ \mathbbm{w} \left( z, q_z+\frac{\chi_i k_i}{2} + \frac{\rho \chi_j k_j}{2} +\frac{n}{\lambda} \right) - \mathbbm{w} \left( z, q_z - \frac{\chi_i k_i}{2} - \frac{\rho \chi_j k_j}{2} - \frac{n}{\lambda} \right) \right],  \notag  
 \end{split} 
\end{align}
\end{widetext}
with $\chi_i = 1$ if $i=1$ and $\chi_i = -1$ if $i=2$.
%
Operators \eqref{eq:FV_Op1} and \eqref{eq:FV_Op2} can be used directly to determine solutions of the system of transport equations for scalar fields \eqref{eq_FVHW1}-\eqref{eq_FVHW2}. 

One major difference between DHW formalism and FVHW formalism becomes again apparent at this stage of the derivation. Within quantum electrodynamics, couplings between fields and particles are based on the Dirac equation, which is first order in spatial derivatives. Consequently, coupling terms are linear in the vector potential and a maximum of one photon per wave can be absorbed per update step. The latter is realized in the form of terms $k_1/2$, $k_2/2$ in the momentum coordinate of the Wigner components \eqref{eqn12b_1}-\eqref{eqn12b_5}.  

This behaviour is entirely different in the Feshbach-Villars formalism. Scalar quantum electrodynamics is based on the Klein-Gordon equation \eqref{equ:FVKG}, thus couplings up to second order may appear. Equations \eqref{eq:FV_Op1} and \eqref{eq:FV_Op2} display this by exhibiting not only a linear (first sum in each of the operators) but also a quadratic term (second set of sums). In this context, at each update step there arises the possibility for a two-photon exchange.

Similarly to the coupling terms in the DHW formalism, the spatial coordinate $z$ appears only within sine and cosine functions. Hence, performing a Fourier transform with respect to $z$ on the complete transport equations \eqref{eq_FVHW1}-\eqref{eq_FVHW2} yields a set of coupled ordinary differential equations \\ 
\begin{alignat}{5}
    &m \partial_t \mathbbm{\tilde f ^+} && && && +2m^2 \mathbbm{\tilde{h}} &&= 0, \\
    &m \partial_t \mathbbm{\tilde f ^-} && -2 P^2 \mathbbm{\tilde h} && +2I \ \mathbbm{\tilde k} && -2m^2 \mathbbm{\tilde{h}} &&= 0, \\  
    \phantom{sum_x^x}
    &m \partial_t \mathbbm{\tilde{h}} && -\phantom{2} P^2 \mathbbm{\tilde f ^+} \hspace{-0.5cm} && && - \phantom{2} m^2 \mathbbm{\tilde f ^+} +m^2 \mathbbm{\tilde f ^-}&&= 0, \\      
    &m \partial_t \mathbbm{\tilde{k}} && && +\phantom{2} I \ \mathbbm{\tilde f ^+} \hspace{-0.5cm} && &&= 0,    
  \end{alignat} 
where we have used the relations $\mathbbm{\tilde f^\pm}=(\mathbbm{\tilde{f}} \pm \mathbbm{\tilde{g}})$. 

The corresponding vacuum initial conditions are transformed accordingly 
\begin{alignat}{3}
  \mathbbm{\tilde f^+}_{\rm vac} \left(K, \boldsymbol{q} \right) = \frac{m}{\sqrt{m^2 + \boldsymbol{q}^2}} \delta \left( K \right), \quad \mathbbm{\tilde{h}}_{\rm vac}  = 0, && \\ 
  \mathbbm{\tilde f^-}_{\rm vac} \left(K, \boldsymbol{q} \right) = \frac{\sqrt{m^2 + \boldsymbol{q}^2}}{m} \delta \left( K \right), \quad \mathbbm{\tilde{k}}_{\rm vac}  = 0. &&  
\end{alignat}
The distribution function then reads
\begin{multline}
n^{FV} \left( K, \boldsymbol{q} \right) = \\
\left( \frac{\sqrt{m^2 + \boldsymbol{q}^2}}{2m} + \frac{K^2}{8m \sqrt{m^2 + \boldsymbol{q}^2}} \right) \Big( \mathbbm{\tilde f^+} - \mathbbm{\tilde f^+}_{\rm vac} \Big) \\
+\frac{m}{2\sqrt{m^2 + \boldsymbol{q}^2}} \Big(\mathbbm{\tilde f^-} - \mathbbm{\tilde f^-}_{\rm vac} \Big).
\end{multline}

The coupling terms in transformed $qK$-space are given by 
\begin{widetext}
\begin{align}
 & - P^2 \mathbbm{w} =  \left( \frac{1}{4} \boldsymbol{D}^2 - \Pi^2 \right) \mathbbm{w} =  \left( -K^2/4 - \boldsymbol{q}^2 \right) \times \left[ \tilde{\mathbbm{w}} \left( K, q_z \right) \right] \label{eq:FV_Op3} \\ 
 \begin{split}
 & \hspace{1cm} +\sum_{i=1}^2 \ \sum_{n=-1}^1 \binom{2}{n+1} \ \sum_{\nu=-1,+1} \ \frac{e}{8} 
 \left( A_{i,x} \left( t \right) \ q_x - \I \nu A_{i,y} \left( t \right) \ q_y \right) \times  \\
 & \hspace{1.5cm} \ee^{\I \nu \omega_i t}
 \left[ \tilde{\mathbbm{w}} \left( K - \nu \left( \chi_i k_i + \frac{2n}{\lambda} \right), q_z + \left( \frac{\chi_i k_i}{2} +\frac{n}{\lambda} \right) \right) \right. \\
 & \hspace{7.5cm} \left. + \tilde{\mathbbm{w}} \left(  K - \nu \left( \chi_i k_i + \frac{2n}{\lambda} \right), q_z - \left( \frac{\chi_i k_i}{2} + \frac{n}{\lambda} \right) \right) \right] \notag
 \end{split} \\
 \begin{split}
 & \hspace{1cm} -\sum_{i=1}^2 \ \sum_{j=1}^2 \ \sum_{\rho=-1,+1} \ \sum_{n=-2}^2 \binom{4}{n+2} \ \sum_{\nu=-1,+1} \ \frac{e^2}{128} \Big( A_{i,x} \left( t \right) A_{j,x} \left( t \right) - \rho A_{i,y} \left( t \right) A_{j,y} \left( t \right) \Big) \times \\
 & \hspace{2cm} \ee^{\I \nu \left( \omega_i + \rho \omega_j \right) t} \left[ \tilde{\mathbbm{w}} \left( K - \nu \left( \chi_i k_i + \rho \chi_j k_j + \frac{2n}{\lambda} \right), q_z + \left( \frac{\chi_i k_i}{2} + \frac{\rho \chi_j k_j}{2} +\frac{n}{\lambda} \right) \right) \right. \\
 & \hspace{4.2cm} + \left. \tilde{\mathbbm{w}} \left( K - \nu \left( \chi_i k_i + \rho \chi_j k_j + \frac{2n}{\lambda} \right), q_z - \left( \frac{\chi_i k_i}{2} + \frac{\rho \chi_j k_j}{2} + \frac{n}{\lambda} \right) \right) \right],  \notag  
 \end{split} 
\end{align}
%
\begin{align}
 & I \mathbbm{w} = \left( \boldsymbol{\Pi} \cdot \boldsymbol{D} \right) \mathbbm{w} =  \left( \I q_z \ K \right) \times \left[ \tilde{\mathbbm{w}} \left( K, q_z \right) \right] \label{eq:FV_Op4} \\
\begin{split}
 & \hspace{1cm} +\sum_{i=1}^2 \ \sum_{n=-1}^1 \binom{2}{n+1} \ \sum_{\nu=-1,+1} \ \frac{e}{8} \left( \I A_{i,x} \nu \left( t \right) q_x + A_{i,y} q_y \left( t \right) \right) \times \\
 & \hspace{1.5cm} \ee^{\I \nu \omega_i t} \left[ \tilde{\mathbbm{w}} \left( K - \nu \left(\chi_i k_i + \frac{2n}{\lambda} \right), q_z+ \left( \frac{\chi_i k_i}{2} +\frac{n}{\lambda} \right) \right) \right. \\
 & \hspace{7.5cm} \left. - \tilde{\mathbbm{w}} \left( K - \nu \left( \chi_i k_i + \frac{2n}{\lambda} \right), q_z - \left( \frac{\chi_i k_i}{2} + \frac{n}{\lambda} \right) \right) \right] \notag 
 \end{split} \\ 
 \begin{split}
 & \hspace{1cm} -\sum_{i=1}^2 \ \sum_{j=1}^2 \ \sum_{\rho=-1,+1} \ \sum_{n=-2}^2 \binom{4}{n+2} \ \sum_{\nu=-1,+1} \ \frac{\I e^2 \nu}{128} \Big( A_{i,x} \left( t \right) A_{j,x} \left( t \right) - \rho A_{i,y} \left( t \right) A_{j,y} \left( t \right) \Big) \times \\
 & \hspace{2cm} \ee^{\I \nu \left( \omega_i + \rho \omega_j \right) t} \left[ \tilde{\mathbbm{w}} \left( K - \nu \left( \chi_i k_i + \rho \chi_j k_j + \frac{2n}{\lambda} \right), q_z + \left( \frac{\chi_i k_i}{2} + \frac{\rho \chi_j k_j}{2} +\frac{n}{\lambda} \right) \right) \right. \\
 & \hspace{4.2cm} - \left. \tilde{\mathbbm{w}} \left( K - \nu \left( \chi_i k_i + \rho \chi_j k_j + \frac{2n}{\lambda} \right), q_z - \left( \frac{\chi_i k_i}{2} + \frac{\rho \chi_j k_j}{2} + \frac{n}{\lambda} \right) \right) \right].  \notag  
 \end{split}  
\end{align}
\end{widetext}

\section{Conclusions}
\label{sec:Conclusion}

The Heisenberg-Wigner formalism has always been hampered by the curse of dimensionality with its core principle being to fully resolve spatial as well as momentum coordinates (any problem is given in terms of an at least $2n$ dimensional domain with $n$ being the number of coordinate dependencies). This has, in fact, effectively turned its biggest strength into its biggest weakness. 
The method introduced in this manuscript copes with this issue finally revealing the formalism's full potential.

Over the course of this manuscript we have introduced a class of field configurations that circumvent the dimensionality problem by separating off all momenta where the fields are homogeneous in their respective spatial coordinates. In this context, it is possible to reduce the transport equations to a $1+1$-dimensional domain with the additional two momentum coordinates given as external parameters. As a result, the equations are trivially parallelizable, thus the computational requirements are drastically reduced.

Moreover, we have shown that for a special class of transverse fields the pseudo-differential operators, which are formulated such that in their original form they facilitate momentum derivatives up to infinite order, are easily resolved in terms of Fourier transforms completely eliminating the need for, e.g., operator expansions. In the same vein, integrations over an auxiliary variable connecting the momentum variables in the formalism with the classical kinetic momentum can be carried out analytically further reducing the complexity of the differential operators. 

Additionally, we have discussed applications of the new approach, e.g., in spatially oscillating fields. In such cases, the pseudo-spectral behaviour of the differential operators can be fully eliminated and, thus, these operators are replaced by discrete coupling terms.



\section{Outlook}
\label{sec:Outlook}

The derivations provided in this manuscript yield a computationally improved version of the Heisenberg-Wigner formalism providing the opportunity to explore new areas of the configuration space which have not been accessible so far. 

It should be emphasized, though, that we have only laid the basis for efficiently treating a complete class of problems within phase-space formalisms. In this regard, we have not even touched the subject of computation and numerical feasibility. While the new formulation of the formalism certainly is an improvement, the true strength lies in the fact that new computer code can now be devised incorporating the conceptual differences at a fundamental level, e.g., in optimizing particle production rates \cite{PhysRevD.88.045028, DesignerFields, PhysRevA.99.022128, Gelfand-Dikii}. 

\subsection{Beyond single-directional propagation}


The advantage of being able to factor out a specific coordinate persists also for field configurations where two or more transverse fields propagate in different directions. For example, for a scenario where one wave is propagating in direction $z$ and a second wave is propagating in direction $x$ the $y$-coordinate can, when factoring in minimal coupling, still be regarded as an external parameter. While it might still be a colossal undertaking to compute production rates in $4$-dimensional phase-space, fields of the form
\begin{equation}
 \boldsymbol{A} = \begin{pmatrix} 0 \\ A(t,x,z) \\ 0 \end{pmatrix}
 \label{equ:Field2d}
\end{equation}
have the advantage of inheriting many benefits of the systems discussed in this manuscript. One might, for example, adapt the derivative operators given in this article for quasi-$1+1$-dimensional configurations to $2$ dimensions. 

\subsection{Momentum coupling in spatially non-periodic envelopes}

In case of oscillating fields each individual photon carries a specific energy and, thus, momentum which potentially act as tracers to distinguish the different particle pairs \cite{Popov}. In Sec. \ref{sec:Appl} we have shown that for special field configurations the opportunity arises to turn the set of differential equations for a continuous momentum variable $q_z$ into a set of algebraic equations that connect Wigner coefficients only at specific, discrete points.

This procedure has been made particularly easy due to the choice of the envelope function in Eq. \eqref{eq:ATZEnv}. Through implementation of a cosine-squared function the Fourier transforms in the vector potential can be carried out analytically. Moreover, the results are given in terms of delta functions which eliminate the subsequent convolution integrals.

If a different envelope function is chosen, evaluating the derivative operators is more difficult. Nevertheless, we might still be able to find approximations that reproduce the full solution accurately even for spatially non-periodic interaction regions.

There are multiple ways to proceed depending on the form of the envelope function. For example, we want to assume that the functions $A_{Z,x} \left( z \right)$ and $A_{Z,y} \left( z \right)$, c.f. Eq. \eqref{eq:ATZEnv}, are only slowly varying in $z$ compared to the underlying oscillating functions \cite{HeinzlKing}. Hence, the impact of the envelope on the momentum distribution is expected to be small. Consequently, we should be able to safely perform a Taylor expansion with respect to $z$ in order to obtain 
\begin{align}
 &A_{Z,x} \left( z \pm \frac{s_z}{2} \right) = \label{eq:ATZxE} \\
 & && \hspace{-1.5cm} A_{Z,x} \left( z \right) \pm \frac{s_z}{2} \partial_z A_{Z,x} \left( z \right) + \mathcal{O} \approx A_{Z,x} \left( z \right), \notag \\
 &A_{Z,y} \left( z \pm \frac{s_z}{2} \right) = \label{eq:ATZyE} \\
 & && \hspace{-1.5cm} A_{Z,y} \left( z \right) \pm \frac{s_z}{2} \partial_z A_{Z,y} \left( z \right) + \mathcal{O} \approx A_{Z,y} \left( z \right). \notag
\end{align}
In this way, the envelope function is unaffected by a Fourier transform in $s_z$ but still restricts the interaction region to a finite volume.


\subsection{Energy-Momentum channels}

A transformation from coordinate space to energy-momentum channels as performed in Sec. \ref{sec:Appl} is not advised for general fields. The reason is that, while a transformation of the envelope functions might yield a simple analytical result, the following convolution with the Wigner coefficient only shows a closed analytical expression under very specific circumstances, for example, if the envelope can be written in terms of an oscillating function. 

It should be further stressed that for setups with two different oscillation frequencies paraphrasing the transport equations in terms of discrete channel equations might even be unfeasible. This is due to the fact that information between these channels is exchanged in steps of $k_1$ and $k_2$. While this is clearly not a problem for, e.g., scattering of a pulse beam with a frequency-doubled probe beam, for setups with wildly irregular frequencies such a flow of information could amount to an absurdly high number of channels to consider. 

Nevertheless, formulating pair production within the Heisenberg-Wigner formalism in terms of coupled discrete channel equations remains an interesting concept, especially when considering the success of Furry-picture quantization \cite{Aleksandrov:2016lxd, fradkin_gitman_shvartsman, Furry:1951zz}.

\subsection{Beyond propagating wave configurations}

The exemplary field configuration discussed in this work capture by no means all possible setups the formalism is capable of describing. The toy model displayed only serves as a reference point for future studies. Such computations might include localized, non-propagating fields \cite{Hebenstreit} as well as chirped laser pulses \cite{PhysRevD.82.045007,PhysRevD.104.016009}.

\subsection{Non-abelian quantum plasmas and chiral kinetic theory}

The derivations in this manuscript have been performed having light-by-light interactions in mind. However, the Heisenberg-Wigner formalism is universally applicable and is therefore already used in various branches of physics, c.f. quantum chromodynamics and quark-gluon transport theory \cite{Elze:1989un, Elze:1986hq} or chiral kinetic theory \cite{Wang:2019moi, Sheng:2017lfu, PhysRevResearch.2.023257}.

As the resulting transport equations are all structurally very similar, the findings given in this manuscript might also turn out to be useful in other areas. 

\section{Acknowledgments}

We are grateful to Ralf Sch\"utzhold and Ivan Aleksandrov for the discussions that inspired this work. We thank the Institute for Theoretical Physics at Kyoto University, because discussions during the workshop ``The Schwinger Effect and Strong-Field Physics'' were useful to complete this work. \\


\appendix

\section{Lower dimensional systems}

The derivation of transport equations for the Wigner components in other than three dimensions works by the same principles as detailed in Secs.\ref{sec:Appl_DHW} and \ref{sec:Appl_FVHW}.

The most notable difference is the change in basis matrices and, therefore, a change in the characteristics of particle fields. In the case of the DHW formalism in $2+1$ dimensions, for example, the problem arises that the Lagrangian can be posed in different ways (reducible or irreducible representation) \cite{KohlfurstDiss}. 

Nevertheless, as the basic principles of the formalism do not change it is still possible to derive a lower-dimensional system from a higher-dimensional set of transport equations. This might prove useful, because in case of transverse fields of the form Eq. \eqref{equ:Field} the resulting set of differential equations shows decreased complexity ($p_y=0$). For the sake of completeness, we will derive the corresponding $2+1$-dimensional transport equations in the following.


\subsection{DHW formalism in 2+1 dimensions}

In $2+1$ dimensions, this is when we disregard the $y$-direction (on the basis of employing fields of the form Eq. \eqref{equ:Field}), the system of equations \eqref{eq_DHW1}-\eqref{eq_DHW2} for the DHW formalism automatically simplifies to
\begin{widetext}
\begin{alignat}{8}
    & \partial_t \mathbbm{s} && && && -2 q_x \mathbbm{t}_\mathbbm{1} {}_{,1} && && -2 q_z \mathbbm{t}_\mathbbm{1} {}_{,3}  && -2 \Pi_1 \mathbbm{t}_\mathbbm{1} {}_{,1} &&= 0, \label{eq_8_1} \\     
    & \partial_t \mathbbm{v}_\mathbb{0} &&+ D_1 \mathbbm{v}_1 && + \partial_z \mathbbm{v}_3 && && && && &&= 0,  \\
    & \partial_t \mathbbm{v}_1 && + D_1 \mathbbm{v}_\mathbb{0} && && && && -2 q_z \mathbbm{a}_{2}  && &&= -2m \mathbbm{t}_\mathbbm{1} {}_{,1} ,  \\   
    & \partial_t \mathbbm{v}_3 && && + \partial_z \mathbbm{v}_\mathbb{0} && +2 q_x \mathbbm{a}_{2} && && && +2 \Pi_1 \mathbbm{a}_2 &&= -2m\mathbbm{t}_\mathbbm{1} {}_{,3} ,  \\ 
    & \partial_t \mathbbm{a}_2 && && && -2 q_x \mathbbm{v}_{3} && && +2 q_z \mathbbm{v}_{1} && -2 \Pi_1 \mathbbm{v}_3 &&= 0,  \\ 
    & \partial_t \mathbbm{t}_\mathbbm{1} {}_{,1}  && && - \partial_z \mathbbm{t}_\mathbbm{2} {}_{,2} && + 2 q_x \mathbbm{s} && && && +2 \Pi_1 \mathbbm{s} &&= +2m\mathbbm{v}_1, \\ 
    & \partial_t \mathbbm{t}_\mathbbm{1} {}_{,3}  && + D_1 \mathbbm{t}_\mathbbm{2} {}_{,2} && && && && +2 q_z \mathbbm{s} && &&= +2m\mathbbm{v}_3 \\    
    & \partial_t \mathbbm{t}_\mathbbm{2} {}_{,2} && + D_1 \mathbbm{t}_\mathbbm{1} {}_{,3} && - \partial_z \mathbbm{t}_\mathbbm{1} {}_{,1} && && && && &&= 0. \label{eq_8_2}
\end{alignat}
\end{widetext}
By disregarding the third dimension implying $p_y=0$, there cannot be any particle creation or even propagation outside the $xz$-plane. Consequently, all derivatives with respect to the $y$-component vanish. In turn, the system of transport equations splits into two subsystems with the first system describing particle production in external fields \eqref{eq_8_1}-\eqref{eq_8_2}. The second subsystem is equally zero at all times.

To provide an explanation for this decoupling consider that while in three spatial dimensions the intrinsic particle spin is described by a vector field, in two spatial dimensions the particle spin is reduced to a scalar quantity. Consequently, two of the vector components have to vanish. 

As a matter of fact, the set of equations \eqref{eq_8_1}-\eqref{eq_8_2} is exactly the system one would obtain if one started with a reducible representation in $2+1$ dimensions, see Ref. \cite{KohlfurstDiss}. A specialty of an odd number of space-time dimensions, however, is that there are multiple ways to formulate the underlying field theory. The Heisenberg-Wigner equations are no exception. Employing an irreducible representation instead of the basis used to obtain Eqs. \eqref{eq_8_1}-\eqref{eq_8_2} yields a related, but again different, set of transport equations. 

In order to obtain the equations of motion in an irreducible representation, we introduce a mapping of the form 
\begin{alignat}{6}
   &\begin{pmatrix} {}^+\mathbbm{s} \\ {}^-\mathbbm{s} \end{pmatrix} && = 
   \begin{pmatrix} 1/2 & 1/2 \\ 1/2 & -1/2 \end{pmatrix}
   \begin{pmatrix} \mathbbm{s} \\ \mathbbm{a}_2 \end{pmatrix}, \\
   & && \begin{pmatrix} {}^+\mathbbm{v}_1 \\ {}^-\mathbbm{v}_1 \end{pmatrix} = 
   \begin{pmatrix} 1/2 & -1/2 \\ 1/2 & 1/2 \end{pmatrix}
   \begin{pmatrix} \mathbbm{v}_1 \\ \mathbbm{t}_\mathbbm{1} {}_{,3} \end{pmatrix}, \\ 
   &\begin{pmatrix} {}^+\mathbbm{v}_3 \\ {}^-\mathbbm{v}_3 \end{pmatrix} && = 
   \begin{pmatrix} 1/2 & 1/2 \\ 1/2 & -1/2 \end{pmatrix}
   \begin{pmatrix} \mathbbm{v}_3 \\ \mathbbm{t}_\mathbbm{1} {}_{,1} \end{pmatrix}, \\
   & && \begin{pmatrix} {}^+\mathbbm{v}_0 \\ {}^-\mathbbm{v}_0 \end{pmatrix} = 
   \begin{pmatrix} 1/2 & -1/2 \\ 1/2 & 1/2 \end{pmatrix}
   \begin{pmatrix} \mathbbm{v}_0 \\ \mathbbm{t}_\mathbbm{2} {}_{,2} \end{pmatrix}.     
\end{alignat}
As already indicated by the superscripts $+$ and $-$, applying such a mapping leads to a decoupling in the system of differential equations \eqref{eq_8_1}-\eqref{eq_8_2}. The two emerging systems of equations precisely coincide with the transport equations for the $2+1$-dimensional Dirac-Heisenberg-Wigner formalism in an irreducible representation \cite{KohlfurstDiss}
\begin{widetext}
\begin{alignat}{8}
    & \partial_t \ {}^+\mathbbm{s} && && && -2 q_x \ {}^+\mathbbm{v}_3 && && +2 q_z \ {}^+\mathbbm{v}_1 && -2 \Pi_1 \ {}^+\mathbbm{v}_3 &&= 0, \label{eq_9_1} \\     
    & \partial_t \ {}^+\mathbbm{v}_\mathbb{0} &&+ D_1 \ {}^+\mathbbm{v}_1 && + \partial_z \ {}^+\mathbbm{v}_3 && && && && &&= 0,  \\
    & \partial_t \ {}^+\mathbbm{v}_1 && + D_1 \ {}^+\mathbbm{v}_\mathbb{0} && && && && -2 q_z \ {}^+\mathbbm{s} && &&= -2m \ {}^+\mathbbm{v}_3,  \\   
    & \partial_t \ {}^+\mathbbm{v}_3 && && + \partial_z \ {}^+\mathbbm{v}_\mathbb{0} && +2 q_x \ {}^+\mathbbm{s} && && && +2 \Pi_1 \ {}^+\mathbbm{s} &&= +2m \ {}^+\mathbbm{v}_1, \label{eq_9_2}
\end{alignat}
%
\begin{alignat}{8}
    & \partial_t \ {}^-\mathbbm{s} && && && +2 q_x \ {}^-\mathbbm{v}_3 && && -2 q_z \ {}^-\mathbbm{v}_1 && +2 \Pi_1 \ {}^-\mathbbm{v}_3 &&= 0, \label{eq_10_1} \\     
    & \partial_t \ {}^-\mathbbm{v}_\mathbb{0} &&+ D_1 \ {}^-\mathbbm{v}_1 && + \partial_z \ {}^-\mathbbm{v}_3 && && && && &&= 0,  \\
    & \partial_t \ {}^-\mathbbm{v}_1 && + D_1 \ {}^-\mathbbm{v}_\mathbb{0} && && && && +2 q_z \ {}^-\mathbbm{s} && &&= +2m \ {}^-\mathbbm{v}_3,  \\   
    & \partial_t \ {}^-\mathbbm{v}_3 && && + \partial_z \ {}^-\mathbbm{v}_\mathbb{0} && -2 q_x \ {}^-\mathbbm{s} && && && -2 \Pi_1 \ {}^-\mathbbm{s} &&= -2m \ {}^-\mathbbm{v}_1. \label{eq_10_2}
\end{alignat}
\end{widetext}
The only non-vanishing Wigner components in the initial vacuum state simultaneously determining initial conditions are given by
\begin{alignat}{7}
& \frac{1}{2}\mathbbm{s}_{\rm vac} \left(q_x, q_z \right) &&= {}^\pm \mathbbm{s}_{\rm vac} \left(q_x, q_z \right) && = -\frac{m}{\sqrt{m^2 +
   q_x^2 + q_z^2}}, \\
& \frac{1}{2}\mathbbm{v}_{1\rm vac} \left( q_x, q_z \right) &&= {}^\pm\mathbbm{v}_{1\rm vac} \left( q_x, q_z \right) && = -\frac{
   q_x}{\sqrt{m^2 + q_x^2 + q_z^2}}, \\
& \frac{1}{2}\mathbbm{v}_{3\rm vac} \left( q_x, q_z \right) &&= {}^\pm\mathbbm{v}_{3\rm vac} \left( q_x, q_z \right) && = -\frac{
   q_z}{\sqrt{m^2 + q_x^2 + q_z^2}}.   
\end{alignat}

The particle distribution function in $2+1$ dimension follows the same principles as the distribution function in $3+1$ dimensions. In the reducible representation we have
\begin{multline}
n \left( z, q_x, q_z \right) = \\
\frac{m \left( \mathbbm{s}-\mathbbm{s}_{\rm vac} \right) + q_x \left( \mathbbm{v}_1-\mathbbm{v}_{1\rm vac} \right) + q_z \left( \mathbbm{v}_3-\mathbbm{v}_{3\rm vac} \right)}{2\sqrt{m^2+q_x^2+q_z^2}} 
\end{multline}
and in the irreducible representation we obtain the relation
\begin{equation}
 n \left( z, q_x, q_z \right) = {}^+n \left( z, q_x, q_z \right) + {}^-n \left( z, q_x, q_z \right),
\end{equation}
with
\begin{multline}
{}^s n \left( z, q_x, q_z \right) = \\
\frac{m \left( {}^s\mathbbm{s}-{}^s\mathbbm{s}_{\rm vac} \right) + q_x \left( {}^s\mathbbm{v}_1-{}^s\mathbbm{v}_{1\rm vac} \right) + q_z \left( {}^s\mathbbm{v}_3-{}^s\mathbbm{v}_{3\rm vac} \right)}{2\sqrt{m^2+q_x^2+q_z^2}},
 \label{equ:n3}
\end{multline}
where the superscript $s=\{+,-\}$ denotes the different systems, Eqs. \eqref{eq_9_1}-\eqref{eq_9_2} or Eqs. \eqref{eq_10_1}-\eqref{eq_10_2}.
The differential operators are independent of the representation
\begin{alignat}{6}
 & D_1 \mathbbm{w} && = \hspace{6.cm} \label{eqn14_1} \\
 & -\I e \ \mathcal{F}^{-1}_{q_z} \left\{ \left[ A_x \left( z+ \frac{s_z}{2},t \right) - A_x \left( z- \frac{s_z}{2},t \right) \right] \mathcal{F}_{q_z} \left\{ \mathbbm{w} \right\} \right\}, \hspace{-10cm} && \notag \\ 
 & \Pi_1 \mathbbm{w} && = \hspace{6.cm} \label{eqn14_2} \\
 &- \frac{e}{2} \ \mathcal{F}^{-1}_{q_z} \left\{ \left[ A_x \left( z+ \frac{s_z}{2},t \right) + A_x \left( z- \frac{s_z}{2},t \right) \right] \mathcal{F}_{q_z} \left\{ \mathbbm{w} \right\} \right\}, \hspace{-10cm} && \notag  
\end{alignat}  
where $\mathbbm{w} = \mathbbm{w} \left(z,\boldsymbol{q},t \right)$ is again the placeholder for any of the Wigner components.

\subsection{FVHW formalism in 2+1 dimensions}

Due to the fact, that the concept of an intrinsic particle spin does not exist for scalar particles no mappings onto smaller spin-related subsystems can be found. Hence, the only improvement that can be achieved in the case of $2+1$ dimensional systems is the trivial reduction in domain size as, e.g., $q_y$ vanishes.

\section{Additional applications}

In the following we will theorize about various additional applications in addition to the configuration discussed in the main text, Sec. \ref{sec:Appl}. Many of the following results can easily be obtained by evaluating the complex field given prior, c.f. Sec. \ref{sec:Appl}, in specific limits. Nevertheless, we will follow a more direct approach in deriving the final coupling terms. In this way, we can discuss specialties of a field type more easily. However, as such an approach brings diminishing returns, we will discuss configurations on the basis of the DHW formalism only.    

\subsection{Locally homogeneous approximation}
\label{App:LHA}

Evaluating the differential operators \eqref{eqn6_1}-\eqref{eqn6_2} is in general a significant computational challenge despite the fact that the integrals from the original formulation have been eliminated. This is due to the fact, that in general a forward Fourier transform and a backwards Fourier transform have to be performed at every single time-step. In the main body of this manuscript we have shown how certain field configurations circumvent this issue by displaying an analytical solution, Sec. \ref{sec:Appl}.

Another way to overcome this problem is by starting with a time-dependent, spatially homogeneous vector potential thus bypassing the need for Fourier transforms altogether. Within the locally homogeneous approximation (also referred to as locale dipole approximation or local density approximation), for example, it is assumed that the spatial variance of the field is only minor and thus the full vector potential can be replaced by a first-order Taylor expansion
\begin{equation}
 A_x \left( z \pm \frac{s_z}{2},t \right) \approx A_x \left( z,t \right), \quad A_y \left( z \pm \frac{s_z}{2},t \right) \approx A_y \left( z,t \right).
\end{equation}
Consequently, within this approximation the vector potentials are independent of the relative coordinate $s_z$, therefore we can make use of the identity 
\begin{equation}
 \mathcal{F}^{-1}_{q_z} \ \left\{ A(z,t) \ \mathcal{F}_{q_z} \left\{ \mathbbm{w} \right\} \right\} = A(z,t) \ \mathbbm{w}. 
\end{equation}
Additionally, if the particle density can be evaluated independently at any given point in coordinate space $z$, as proposed by the locally homogeneous approximation, derivatives of the Wigner components with respect to $z$ vanish, too. In conclusion, the governing equations of motion simplify dramatically losing all non-local quantities and therefore turning into a set of ordinary differential equations. 

In this way, from the $16$ original Wigner components for the DHW formalism only $10$ are present in the final set of equations. Most notably, the charge density $\mathbbm{v}_\mathbb{0}$ and magnetic moment density $\boldsymbol{\mathbbm{t}}_\mathbbm{2}$ vanish as electrons and positrons cannot be spatially separated any more.  


\subsection{Assisting Potential}
\label{App:Assist}

Another interesting configuration is given by a pulse approaching a strong, slowly varying or even spatially purely homogeneous field, c.f. the dynamical assistance effect \cite{PhysRevLett.101.130404, aleksandrov_prd_2018, Jansen:2013dea, torgrimsson_prd_2018, PhysRevD.100.116018}. Catalysis mechanism for non-perturbative vacuum electron-positron pair production can be approximated by such a configuration \cite{Dunne:2009gi}.

The vector potential is given by 
\begin{align}
 & A_x (z \pm \frac{s_z}{2},t) = A_{s,x}(t) + A_{f,x} \cos \left( \omega t + k \left( z \pm \frac{s_z}{2} \right) \right), \\
 & A_y (z \pm \frac{s_z}{2},t) = A_{s,y}(t) + A_{f,y} \sin \left( \omega t + k \left( z \pm \frac{s_z}{2} \right) \right),
\end{align}
with a slowly varying potential $A_{s} (t)$ and a fast wave $A_{f} \left( \omega t + k z \right)$ propagating with energy $\omega$ and momentum $k$. The core of the operators \eqref{eqn6_1}-\eqref{eqn6_2} thus takes the form
\begin{alignat}{5}
& \left[ A_x \left( z+ \frac{s_z}{2},t \right) - A_x \left( z- \frac{s_z}{2},t \right) \right] && = \notag \\
& && \hspace{-3cm} -2 A_{f,x} \ \sin \left( \omega t + k z \right) \sin \left( \frac{k s_z}{2} \right), \label{eq:AfAs1} && \\
& \left[ A_y \left( z+ \frac{s_z}{2},t \right) - A_y \left( z- \frac{s_z}{2},t \right) \right] && = \notag \\
& && \hspace{-3cm} +2 A_{f,y} \ \cos \left( \omega t + k z \right) \sin \left( \frac{k s_z}{2} \right), \label{eq:AfAs2} \\
& \left[ A_x \left( z+ \frac{s_z}{2},t \right) + A_x \left( z- \frac{s_z}{2},t \right) \right] && = \notag \\
& \hspace{1cm} 2 A_{s,x} && \hspace{-3cm} + 2 A_{f,x} \ \cos \left( \omega t + k z \right) \cos \left( \frac{k s_z}{2} \right). \label{eq:AfAs3} \\
& \left[ A_y \left( z+ \frac{s_z}{2},t \right) + A_y \left( z- \frac{s_z}{2},t \right) \right] && = \notag \\
& \hspace{1cm} 2 A_{s,y} && \hspace{-3cm} + 2 A_{f,y} \ \sin \left( \omega t + k z \right) \cos \left( \frac{k s_z}{2} \right). \label{eq:AfAs4}
\end{alignat}
Plugging in Eqs. \eqref{eq:AfAs1}-\eqref{eq:AfAs4} into the definitions of the original differential operators \eqref{eqn6_1}-\eqref{eqn6_2} and performing the inverse Fourier transform yields 
\begin{alignat}{6}
 & D_1 \mathbbm{w} && = && -e A_{f,x} \sin \left(\omega t + k z \right) \label{eqn9_1} \\ 
 & && \times \left[ \mathbbm{w} \left( z, q_z+\frac{k}{2} \right) - \mathbbm{w} \left( z, q_z - \frac{k}{2} \right) \right], \hspace{-6cm} && \hspace{6cm} \notag \\ 
 & D_2 \mathbbm{w} && = && +e A_{f,y} \cos \left(\omega t + k z \right) \\
 & && \times \left[ \mathbbm{w} \left( z, q_z+\frac{k}{2} \right) - \mathbbm{w} \left( z, q_z - \frac{k}{2} \right) \right], \hspace{-6cm} && \hspace{6cm} \notag \\
 & \Pi_1 \mathbbm{w} && = -e A_{s,x} &&- \frac{e}{2} A_{f,x} \cos \left(\omega t + k z \right) \\
 & && \times \left[ \mathbbm{w} \left( z, q_z- \frac{k}{2} \right) + \mathbbm{w} \left( z, q_z+ \frac{k}{2} \right) \right], \hspace{-6cm} && \hspace{6cm} \notag \\ 
 & \Pi_2 \mathbbm{w} && = -e A_{s,y} &&- \frac{e}{2} A_{f,y} \sin \left(\omega t + k z \right) \label{eqn9_3} \\
 & && \times \left[ \mathbbm{w} \left( z, q_z- \frac{k}{2} \right) + \mathbbm{w} \left( z, q_z+ \frac{k}{2} \right) \right], \hspace{-6cm} && \hspace{6cm} \notag 
\end{alignat}  
and, additionally, 
\begin{equation}
D_3 = \partial_z \quad {\text{and}} \quad \Pi_3= 0.    
\label{eqn9_2}
\end{equation}

At this point we again observe that the coordinate $z$ only shows up in the form $\sin \left( kz \right)$, $\cos \left( kz \right)$ and $\partial_z$, leaving us with the option to pose the system not in coordinate-space, but in terms of absorption channels. Expanding the Wigner coefficients in a Fourier series
\begin{equation}
 \mathbbm{w} \left(z, \boldsymbol{q}, t \right) = \sum_j \ \mathbbm{w}^j \left(\boldsymbol{q}, t \right) \ e^{\I K_j z},
\end{equation}
with the momentum channels $K_j=jk$, completely restructures the system of transport equation and, with it, the operators in Eqs. \eqref{eqn9_1}-\eqref{eqn9_2}. The coupling terms are given by
\begin{widetext}
\begin{align}
 \begin{split}
 D_1 \mathbbm{w} & = \hspace{2.5cm} + \frac{\I e A_{f,x}}{2} \times 
   \left\{ e^{+\I \omega t} \left[ \mathbbm{w}^{j-1} \left( q_z+\frac{k}{2} \right) - \mathbbm{w}^{j-1} \left( q_z - \frac{k}{2} \right) \right] \right. \\ 
  & \hspace{4.8cm} - \left. e^{-\I \omega t} \left[ \mathbbm{w}^{j+1} \left( q_z+\frac{k}{2} \right) - \mathbbm{w}^{j+1} \left( q_z - \frac{k}{2} \right) \right] \right\}, \label{eqn10_1} \\ 
 \end{split} \\
 \begin{split}
 D_2 \mathbbm{w} & = \hspace{2.6cm} + \frac{e A_{f,y}}{2} \times 
   \left\{ e^{+\I \omega t} \left[ \mathbbm{w}^{j-1} \left( q_z+\frac{k}{2} \right) - \mathbbm{w}^{j-1} \left( q_z - \frac{k}{2} \right) \right] \right. \\ 
  & \hspace{4.8cm} + \left. e^{-\I \omega t} \left[ \mathbbm{w}^{j+1} \left( q_z+\frac{k}{2} \right) - \mathbbm{w}^{j+1} \left( q_z - \frac{k}{2} \right) \right] \right\}, \label{eqn10_2} \\ 
 \end{split} \\ 
 \begin{split}
 \Pi_1 \mathbbm{w} & = -e A_{s,x} \mathbbm{w}^{j} \left( q_z \right) - \frac{e A_{f,x}}{4} \times 
   \left\{ e^{+\I \omega t} \left[ \mathbbm{w}^{j-1} \left( q_z+ \frac{k}{2} \right) + \mathbbm{w}^{j-1} \left( q_z- \frac{k}{2} \right) \right] \right. \\
   & \hspace{4.9cm} + \left. e^{-\I \omega t} \left[ \mathbbm{w}^{j+1} \left( q_z+ \frac{k}{2} \right) + \mathbbm{w}^{j+1} \left( q_z- \frac{k}{2} \right) \right] \right\}, \label{eqn10_4} \\  
 \end{split} \\
 \begin{split}
 \Pi_2 \mathbbm{w} & = -e A_{s,y} \mathbbm{w}^{j} \left( q_z \right) + \frac{\I e A_{f,y}}{4} \times 
   \left\{ e^{+\I \omega t} \left[ \mathbbm{w}^{j-1} \left( q_z+ \frac{k}{2} \right) + \mathbbm{w}^{j-1} \left( q_z- \frac{k}{2} \right) \right] \right. \\
   & \hspace{4.9cm} - \left. e^{-\I \omega t} \left[ \mathbbm{w}^{j+1} \left( q_z+ \frac{k}{2} \right) + \mathbbm{w}^{j+1} \left( q_z- \frac{k}{2} \right) \right] \right\}. \label{eqn10_5} \\  
 \end{split} \\
\end{align}  
\end{widetext}
In direction of wave propagation we obtain
\begin{equation}
D_3 = \I \left(j k \right) \mathbbm{w}^j \left(q_z \right) \quad {\text{and}} \quad \Pi_3= 0.    
\label{eqn10_6}  
\end{equation}


Vacuum initial conditions as well as the definitions for the particle distribution function and, thus, the particle spectrum have to be changed accordingly.

\subsection{Standing Waves}
\label{App:Stand}

In the collision of two identical, counter-propagating laser beams a standing wave pattern is created \cite{Ringwald:2001ib, Ribeyre, PhysRevLett.87.193902, Aleksandrov:2017mtq}. This idealized scenario does not take into account the spatial finiteness of the beams. Nevertheless, through proper modeling of the interaction region accurate predictions regarding the kinematics of the particles created can be obtained \cite{KohlfurstMass, KohlfurstDirac, PhysRevD.91.125026}.   

The corresponding vector potential is given by a toy model of the form
\begin{align}
 &A_x \left( z \pm \frac{s_z}{2},t \right) = A_1(t) \cos \left( k \left( z \pm \frac{s_z}{2} \right) \right), \\
 &A_y \left( z \pm \frac{s_z}{2},t \right) = A_2(t) \cos \left( k \left( z \pm \frac{s_z}{2} \right) \right). \label{eq:Acos}
\end{align}
Depending on the choice of $A_1(t)$ and $A_2(t)$ the polarization of the beam can be adjusted. Furthermore, if one is only interested in the region in the vicinity of one of the interaction region's peaks the local density approximation (locally homogeneous approximation, see Appendix \ref{App:LHA}) can be applied significantly reducing the problem's complexity,
\begin{alignat}{5}
& \left[ A_x \left( z+ \frac{s_z}{2},t \right) - A_x \left( z- \frac{s_z}{2},t \right) \right] && = \\
& && \hspace{-2cm} -2 A_1(t) \sin \left(k z \right) \sin \left( \frac{k s_z}{2} \right), \notag \\
& \left[ A_y \left( z+ \frac{s_z}{2},t \right) - A_y \left( z- \frac{s_z}{2},t \right) \right] && = \\
& && \hspace{-2cm} -2 A_2(t) \sin \left(k z \right) \sin \left( \frac{k s_z}{2} \right), \notag \\
& \left[ A_x \left( z+ \frac{s_z}{2},t \right) + A_x \left( z- \frac{s_z}{2},t \right) \right] && = \\
& && \hspace{-2cm} +2 A_1(t) \cos \left(k z \right) \cos \left( \frac{k s_z}{2} \right), \notag \\
& \left[ A_y \left( z+ \frac{s_z}{2},t \right) + A_y \left( z- \frac{s_z}{2},t \right) \right] && = \\
& && \hspace{-2cm} +2 A_2(t) \cos \left(k z \right) \cos \left( \frac{k s_z}{2} \right). \notag
\end{alignat}
By performing Fourier transforms we obtain operators of the form
\begin{alignat}{6}
 & D_1 \mathbbm{w} && = && -e A_1(t) \sin \left(k z \right) \label{eqn7_1} \\
 & && \times \left[ \mathbbm{w} \left( z, q_z+\frac{k}{2} \right) - \mathbbm{w} \left( z, q_z - \frac{k}{2} \right) \right], \hspace{-6cm} && \hspace{6cm} \notag \\ 
 & D_2 \mathbbm{w} && = && -e A_2(t) \sin \left(k z \right) \\
 & && \times \left[ \mathbbm{w} \left( z, q_z+\frac{k}{2} \right) - \mathbbm{w} \left( z, q_z - \frac{k}{2} \right) \right], \hspace{-6cm} && \hspace{6cm} \notag \\
 & \Pi_1 \mathbbm{w} && = &&- \frac{e}{2} A_1(t) \cos \left(k z \right) \\
 & && \times \left[ \mathbbm{w} \left( z, q_z+ \frac{k}{2} \right) + \mathbbm{w} \left( z, q_z- \frac{k}{2} \right) \right], \hspace{-6cm} && \hspace{6cm} \notag \\ 
 & \Pi_2 \mathbbm{w} && = &&- \frac{e}{2} A_2(t) \cos \left(k z \right) \\
 & && \times \left[ \mathbbm{w} \left( z, q_z+ \frac{k}{2} \right) + \mathbbm{w} \left( z, q_z- \frac{k}{2} \right) \right], \hspace{-6cm} && \hspace{6cm} \notag  
\end{alignat}  
and, additionally,
\begin{equation}
 D_3 \mathbbm{w} = \partial_z \mathbbm{w} \quad \text{and} \quad \Pi_3 = 0.   \label{eqn7_2}
\end{equation}

As in the main body of the manuscript, instead of performing derivatives for a continuous parameter $q_z$, the momentum in $z$-direction is now treated as a discrete variable. By introducing a vector potential of the form of Eq. \eqref{eq:Acos} and by demanding periodic boundary conditions in $z$ we have restricted the quantum system in such a way that particle-photon interactions can only occur through absorption/emission of quanta of size $k$. Moreover, as we are still operating in center-of-momentum coordinates an increase in particle momentum of $k$ shifts the coordinate system by $k/2$ thus preserving symmetry.

For the sake of completeness, an expansion in terms of Fourier coefficients
\begin{equation}
 \mathbbm{w} \left(z, \boldsymbol{q}, t \right) = \sum_j \ \mathbbm{w}^j \left(\boldsymbol{q}, t \right) \ e^{\I K_j z},
\end{equation}
with the momentum channels $K_j=jk$ is in order. Consequently, we obtain coupled channel equations employing a coupling procedure determined by the operators
\begin{widetext}
\begin{alignat}{8}
 & D_1 \mathbbm{w} && = && - \frac{\I e A_1(t)}{2} \times && 
 & && && \left[ \mathbbm{w}^{j-1} \left( q_z-\frac{k}{2} \right) - \mathbbm{w}^{j-1} \left( q_z + \frac{k}{2} \right) - \mathbbm{w}^{j+1} \left( q_z-\frac{k}{2} \right) + \mathbbm{w}^{j+1} \left( q_z + \frac{k}{2} \right) \right], && \label{eqn8_1} \\ 
 & D_2 \mathbbm{w} && = && - \frac{\I e A_2(t)}{2} \times && 
 & && && \left[ \mathbbm{w}^{j-1} \left( q_z-\frac{k}{2} \right) - \mathbbm{w}^{j-1} \left( q_z + \frac{k}{2} \right) - \mathbbm{w}^{j+1} \left( q_z-\frac{k}{2} \right) + \mathbbm{w}^{j+1} \left( q_z + \frac{k}{2} \right) \right], && \\
 & \Pi_1 \mathbbm{w} && = && - \frac{e A_1(t)}{4} \times && 
 & && && \left[ \mathbbm{w}^{j-1} \left( q_z- \frac{k}{2} \right) + \mathbbm{w}^{j-1} \left( q_z+ \frac{k}{2} \right) + \mathbbm{w}^{j+1} \left( q_z- \frac{k}{2} \right) + \mathbbm{w}^{j+1} \left( q_z+ \frac{k}{2} \right) \right], &&  \\ 
 & \Pi_2 \mathbbm{w} && = &&- \frac{e A_2(t)}{4} \times &&
 & && && \left[ \mathbbm{w}^{j-1} \left( q_z- \frac{k}{2} \right) + \mathbbm{w}^{j-1} \left( q_z+ \frac{k}{2} \right) + \mathbbm{w}^{j+1} \left( q_z- \frac{k}{2} \right) + \mathbbm{w}^{j+1} \left( q_z+ \frac{k}{2} \right) \right], &&  
\end{alignat}  
\end{widetext}
as well as 
\begin{equation}
 D_3 \mathbbm{w} = + \I \left(j k \right) \mathbbm{w}^j \left(q_z \right), \quad \text{and} \quad \Pi_3 = 0.   \label{eqn8_2}
\end{equation}
The overall structure is very similar to the system of operators developed previously. The only difference is that the field frequency plays a slightly different role here. For potentials with a second, homogeneous potential the frequency of the fast field serves as a weighting factor for the various coefficients, see Appendix \ref{App:Assist}. For a standing wave, however, both waves exhibit the same fundamental oscillation frequency thus there are no weighting factors and the time-dependent part of the potentials is relegated to a prefactor.

Vacuum initial conditions as well as the particle distribution function have to be adjusted accordingly.

\subsection{Bi-frequent Fields}
\label{App:Bi}

In scenarios with two counterpropagating electromagnetic waves without a clear separation between slow and fast fields, both field frequencies are of equal importance \cite{Panferov:2015yda, Akal:2014eua, Otto:2014ssa}. Hence, such a configuration automatically doubles the amount of components in the differential or coupling operators.
Examples of such a set-up are also given by, e.g., light-by-light scattering of a pump laser pulse and a frequency-doubled probe beam \cite{KohlfurstLight}.

A typical field configuration to model such a case is given by
\begin{align}
 \begin{split}
 A_x (z \pm \frac{s_z}{2},t) &= \\
 & A_{1,x} \left( t \right) \ \cos \left( \omega_1 t + k_1 \left( z \pm \frac{s_z}{2} \right) \right)  \\
 &  + A_{2,x} \left( t \right) \ \cos \left( \omega_2 t - k_2 \left( z \pm \frac{s_z}{2} \right) \right) , \label{eq:ATZx}
 \end{split} \\
 \begin{split}
 A_y (z \pm \frac{s_z}{2},t) &= \\
  & A_{1,y} \left( t \right) \ \sin \left( \omega_1 t + k_1 \left( z \pm \frac{s_z}{2} \right) \right) \\
  &  + A_{2,y} \left( t \right) \ \sin \left( \omega_2 t - k_2 \left( z \pm \frac{s_z}{2} \right) \right) . \label{eq:ATZy}
 \end{split} 
\end{align}
where $\omega_1$, $\omega_2$ determine the energy of the individual photons in the beam and $k_1$, $k_2$ give their linear momentum, respectively. To ensure that the vector potential vanishes at asymptotic times we have further introduced the envelope functions $A_{1,x} \left( t \right)$, $A_{2,x} \left( t \right)$, $A_{1,y} \left( t \right)$ and $A_{2,y} \left( t \right)$. These scalar functions also serve the purpose of defining the polarization of the two electromagnetic waves. 

In this setup, the potential-dependent factors within the differential operators \eqref{eqn6_1}-\eqref{eqn6_2} become
\begin{align}
 \begin{split}
 \left[ A_x \right. & \left. \left( z+ \frac{s_z}{2},t \right) - A_x \left( z- \frac{s_z}{2},t \right) \right] = \\
 & - 2 \left( A_{1,x} \left( t \right) \ \sin \left( \omega_1 t + k_1 z \right) \sin \left( \frac{k_1 s_z}{2} \right) \right. \\
 & \hspace{0.7cm} \left. - A_{2,x} \left( t \right) \ \sin \left( \omega_2 t - k_2 z \right) \sin \left( \frac{k_2 s_z}{2} \right) \right), \label{eq:A1A21}
\end{split} \\
\begin{split}
 \left[ A_y \right. & \left. \left( z+ \frac{s_z}{2},t \right) - A_y \left( z- \frac{s_z}{2},t \right) \right] = \\
 & + 2 \left( A_{1,y} \left( t \right) \ \cos \left( \omega_1 t + k_1 z \right) \sin \left( \frac{k_1 s_z}{2} \right) \right. \\
 & \hspace{0.7cm} \left. - A_{2,y} \left( t \right) \ \cos \left( \omega_2 t - k_2 z \right) \sin \left( \frac{k_2 s_z}{2} \right) \right), \label{eq:A1A22} 
\end{split} 
\end{align}
\begin{align}
 \begin{split}
 \left[ A_x \right. & \left. \left( z+ \frac{s_z}{2},t \right) + A_x \left( z- \frac{s_z}{2},t \right) \right] = \\
 & + 2 \left( A_{1,x} \left( t \right) \ \cos \left( \omega_1 t + k_1 z \right) \cos \left( \frac{k_1 s_z}{2} \right) \right. \\
 & \hspace{0.7cm} \left. + A_{2,x} \left( t \right) \ \cos \left( \omega_2 t - k_2 z \right) \cos \left( \frac{k_2 s_z}{2} \right) \right), \label{eq:A1A23}
 \end{split} \\
 \begin{split}
 \left[ A_y \right. & \left. \left( z+ \frac{s_z}{2},t \right) + A_y \left( z- \frac{s_z}{2},t \right) \right] = \\
 &  + 2 \left( A_{1,y} \left( t \right) \ \sin \left( \omega_1 t + k_1 z \right) \cos \left( \frac{k_1 s_z}{2} \right) \right. \\
 & \hspace{0.7cm} \left. + A_{2,y} \left( t \right) \ \sin \left( \omega_2 t - k_2 z \right) \cos \left( \frac{k_2 s_z}{2} \right) \right). \label{eq:A1A24}
 \end{split} 
\end{align} 

Accordingly, after performing the inverse Fourier transform the differential operators take on the form 
\begin{widetext}
\begin{alignat}{6}
 & D_1 \mathbbm{w} && = && -e  A_{1,x} \left( t \right) \sin \left(\omega_1 t + k_1 z \right) \ && \left[ \mathbbm{w} \left( z, q_z+\frac{k_1}{2} \right) - \mathbbm{w} \left( z, q_z - \frac{k_1}{2} \right) \right] && \notag \\
 & && && +e  A_{2,x} \left( t \right) \sin \left(\omega_2 t - k_2 z \right) \ && \left[ \mathbbm{w} \left( z, q_z+\frac{k_2}{2} \right) - \mathbbm{w} \left( z, q_z - \frac{k_2}{2} \right) \right], && \label{eqn11_1} \\ 
 & D_2 \mathbbm{w} && = && +e  A_{1,y} \left( t \right) \cos \left(\omega_1 t + k_1 z \right) \ && \left[ \mathbbm{w} \left( z, q_z+\frac{k_1}{2} \right) - \mathbbm{w} \left( z, q_z - \frac{k_1}{2} \right) \right] && \notag \\
 & && && -e  A_{2,y} \left( t \right) \cos \left(\omega_2 t - k_2 z \right) \ && \left[ \mathbbm{w} \left( z, q_z+\frac{k_2}{2} \right) - \mathbbm{w} \left( z, q_z - \frac{k_2}{2} \right) \right], && \label{eqn11_2} \\  
 & \Pi_1 \mathbbm{w} && = &&- \frac{e}{2}  A_{1,x} \left( t \right) \cos \left(\omega_1 t + k_1 z \right) \ && \left[ \mathbbm{w} \left( z, q_z+ \frac{k_1}{2} \right) + \mathbbm{w} \left( z, q_z- \frac{k_1}{2} \right) \right] && \notag \\
 & && &&- \frac{e}{2} A_{2,x} \left( t \right) \cos \left(\omega_2 t - k_2 z \right) \ && \left[ \mathbbm{w} \left( z, q_z+ \frac{k_2}{2} \right) + \mathbbm{w} \left( z, q_z- \frac{k_2}{2} \right) \right], && \label{eqn11_4} \\
 & \Pi_2 \mathbbm{w} && = &&- \frac{e}{2}  A_{1,y} \left( t \right) \sin \left(\omega_1 t + k_1 z \right) \ && \left[ \mathbbm{w} \left( z, q_z+ \frac{k_1}{2} \right) + \mathbbm{w} \left( z, q_z- \frac{k_1}{2} \right) \right] && \notag \\
 & && &&- \frac{e}{2} A_{2,y} \left( t \right) \sin \left(\omega_2 t - k_2 z \right) \ && \left[ \mathbbm{w} \left( z, q_z+ \frac{k_2}{2} \right) + \mathbbm{w} \left( z, q_z- \frac{k_2}{2} \right) \right], && \label{eqn11_5} 
\end{alignat}  
\end{widetext}
with
\begin{equation}
 D_3 \mathbbm{w} = \partial_z \mathbbm{w}, \quad \text{and} \quad \Pi_3 = 0.  \label{eqn11_6}
\end{equation}
After converting all derivatives in $q_z$ into simple couplings of Wigner components at different momenta, we clearly see the signatures of both waves in the arguments of the Wigner coefficients. From the arguments we see that with respect to center-of-momentum coordinates a linear momentum transfer can only happen in discrete steps through absorption of a photon with momentum $k_1$ or $k_2$, respectively. 

Again, the spatial dependency is solely given in terms of sine and cosine functions making it exceptionally well suited to treat the system in transformed space; $z \to K$. To show also an alternative representation, however, we give the coupling terms for a continuous variable $K$ instead of discrete energy-momentum channels denoted by $j$. 
   
Note, that neither the transport equations or coupling operators nor the computation procedure changes in any way. At this point such an alternative representation is purely of aesthetic interest. Only for more complex field structures when a simple expansion in terms of clearly separated Fourier modes is not possible any more this ``continuous'' representation reigns supreme, see the operators in Sec. \ref{sec:Appl_DHW} for an arbitrarily sized envelope function,  
\begin{widetext}
\begin{align}
 \begin{split}
 & D_1 \mathbbm{w} = + \frac{\I e A_{1,x} \left( t \right)}{2}  
   \left\{ e^{+\I \omega_1 t} \left[ \tilde{\mathbbm{w}} \left( K - k_1, q_z+\frac{k_1}{2} \right) - \tilde{\mathbbm{w}} \left( K-k_1, q_z - \frac{k_1}{2} \right) \right] \right. \\ 
  & \hspace{3.5cm} - \left. e^{-\I \omega_1 t} \left[ \tilde{\mathbbm{w}} \left( K+k_1, q_z+\frac{k_1}{2} \right) - \tilde{\mathbbm{w}} \left( K+k_1, q_z - \frac{k_1}{2} \right) \right] \right\} \\
  & \hspace{1.3cm} + \frac{\I e A_{2,x} \left( t \right)}{2}  
   \left\{ e^{-\I \omega_2 t} \left[ \tilde{\mathbbm{w}} \left( K - k_2, q_z+\frac{k_2}{2} \right) - \tilde{\mathbbm{w}} \left( K-k_2, q_z - \frac{k_2}{2} \right) \right] \right. \\ 
  & \hspace{3.5cm} - \left. e^{+\I \omega_2 t} \left[ \tilde{\mathbbm{w}} \left( K+k_2, q_z+\frac{k_2}{2} \right) - \tilde{\mathbbm{w}} \left( K+k_2, q_z - \frac{k_2}{2} \right) \right] \right\}, \label{eqn12_1} 
  \raisetag{1.75\normalbaselineskip}
 \end{split} \\
 \begin{split}
 & D_2 \mathbbm{w} = + \frac{e A_{1,y} \left( t \right)}{2}  
   \left\{ e^{+\I \omega_1 t} \left[ \tilde{\mathbbm{w}} \left( K - k_1, q_z+\frac{k_1}{2} \right) - \tilde{\mathbbm{w}} \left( K-k_1, q_z - \frac{k_1}{2} \right) \right] \right. \\ 
  & \hspace{3.5cm} + \left. e^{-\I \omega_1 t} \left[ \tilde{\mathbbm{w}} \left( K+k_1, q_z+\frac{k_1}{2} \right) - \tilde{\mathbbm{w}} \left( K+k_1, q_z - \frac{k_1}{2} \right) \right] \right\} \\
  & \hspace{1.3cm} - \frac{e A_{2,y} \left( t \right)}{2}  
   \left\{ e^{-\I \omega_2 t} \left[ \tilde{\mathbbm{w}} \left( K - k_2, q_z+\frac{k_2}{2} \right) - \tilde{\mathbbm{w}} \left( K-k_2, q_z - \frac{k_2}{2} \right) \right] \right. \\ 
  & \hspace{3.5cm} + \left. e^{+\I \omega_2 t} \left[ \tilde{\mathbbm{w}} \left( K+k_2, q_z+\frac{k_2}{2} \right) - \tilde{\mathbbm{w}} \left( K+k_2, q_z - \frac{k_2}{2} \right) \right] \right\}  , \label{eqn12_2} \raisetag{1.25\normalbaselineskip}
 \end{split} \\ 
 \begin{split} \\
 & \Pi_1 \mathbbm{w} = - \frac{e A_{1,x} \left( t \right)}{4}  
   \left\{ e^{+\I \omega_1 t} \left[ \tilde{\mathbbm{w}} \left( K - k_1, q_z+\frac{k_1}{2} \right) + \tilde{\mathbbm{w}} \left( K-k_1, q_z - \frac{k_1}{2} \right) \right] \right. \\ 
  & \hspace{3.4cm} + \left. e^{-\I \omega_1 t} \left[ \tilde{\mathbbm{w}} \left( K+k_1, q_z+\frac{k_1}{2} \right) + \tilde{\mathbbm{w}} \left( K+k_1, q_z - \frac{k_1}{2} \right) \right] \right\} \\
  & \hspace{1.3cm} - \frac{e A_{2,x} \left( t \right)}{4}  
   \left\{ e^{-\I \omega_2 t} \left[ \tilde{\mathbbm{w}} \left( K - k_2, q_z+\frac{k_2}{2} \right) + \tilde{\mathbbm{w}} \left( K-k_2, q_z - \frac{k_2}{2} \right) \right] \right. \\ 
  & \hspace{3.4cm} + \left. e^{+\I \omega_2 t} \left[ \tilde{\mathbbm{w}} \left( K+k_2, q_z+\frac{k_2}{2} \right) + \tilde{\mathbbm{w}} \left( K+k_2, q_z - \frac{k_2}{2} \right) \right] \right\}, \label{eqn12_4} 
  \raisetag{1.25\normalbaselineskip}
 \end{split} \\
 \begin{split}
 & \Pi_2 \mathbbm{w} = + \frac{\I e A_{1,y} \left( t \right)}{4} 
   \left\{ e^{+\I \omega_1 t} \left[ \tilde{\mathbbm{w}} \left( K - k_1, q_z+\frac{k_1}{2} \right) + \tilde{\mathbbm{w}} \left( K-k_1, q_z - \frac{k_1}{2} \right) \right] \right. \\ 
  & \hspace{3.4cm} - \left. e^{-\I \omega_1 t} \left[ \tilde{\mathbbm{w}} \left( K+k_1, q_z+\frac{k_1}{2} \right) + \tilde{\mathbbm{w}} \left( K+k_1, q_z - \frac{k_1}{2} \right) \right] \right\} \\
  & \hspace{1.3cm} - \frac{\I e A_{2,y} \left( t \right)}{4}  
   \left\{ e^{-\I \omega_2 t} \left[ \tilde{\mathbbm{w}} \left( K - k_2, q_z+\frac{k_2}{2} \right) + \tilde{\mathbbm{w}} \left( K-k_2, q_z - \frac{k_2}{2} \right) \right] \right. \\ 
  & \hspace{3.4cm} - \left. e^{+\I \omega_2 t} \left[ \tilde{\mathbbm{w}} \left( K+k_2, q_z+\frac{k_2}{2} \right) + \tilde{\mathbbm{w}} \left( K+k_2, q_z - \frac{k_2}{2} \right) \right] \right\}, \label{eqn12_5} \raisetag{1.25\normalbaselineskip}  
 \end{split} 
\end{align}  
\end{widetext}
and
\begin{equation}
 D_3 \mathbbm{w} = \left( \I K \right) \tilde{\mathbbm{w}} \left(K, q_z \right), \quad \Pi_3 = 0.  \label{eqn12_6}
\end{equation}

Vacuum initial conditions as well as the definition of the particle distribution have to be adjusted.

\interlinepenalty=10000


\begin{thebibliography}{99}
\bibitem{footnote1}
Derivatives in momentum variables are denoted $\partial^p$, while spatial derivatives are given by $\partial^r$.
\bibitem{footnote2}
Note that here we clearly pursue a formulation in terms of the momentum $\boldsymbol{p}$ instead of the four-momentum $p$. This is due to the fact that we directly derive an equal-time formalism without introducing  
covariant transport equations first. As we consider only initial-value problems there is no need to obtain a covariant formalism anyway.
\bibitem{footnote3}
The multiplicative factors linear in the vector potential are already given in Eqs.  \eqref{eq:A1A21e}-\eqref{eq:A1A24e}. The corresponding quadratic terms only yield lengthy expressions that do not add anything of value if stated in full length. We have thus decided to not write them down explicitly, but instead focus on the transformed operators.
\end{thebibliography}
\end{document}